\documentclass[12pt,preprint]{aastex}		

\shortauthors{Oliveira et al.}
\shorttitle{{\sl Spitzer} Survey of Protoplanetary Disk Dust in Serpens}

\begin{document}

\title{A {\sl Spitzer} Survey of Protoplanetary Disk Dust in the Young
Serpens Cloud: How do Dust Characteristics Evolve with Time?}
\author{ Isa Oliveira\altaffilmark{1,2}, Klaus
  M. Pontoppidan\altaffilmark{2}, Bruno Mer\'{i}n\altaffilmark{3},
  Ewine F. van Dishoeck\altaffilmark{1,4}, Fred
  Lahuis\altaffilmark{5,1}, Vincent C. Geers\altaffilmark{6}, Jes
  K. J{\o}rgensen\altaffilmark{7}, Johan Olofsson\altaffilmark{8},
  Jean-Charles Augereau\altaffilmark{8}, Joanna
  M. Brown\altaffilmark{4}}
\altaffiltext{1}{Leiden Observatory, Leiden University, P.O. Box 9513,
  2300 RA Leiden, The Netherlands, email: oliveira@strw.leidenuniv.nl}
\altaffiltext{2}{California Institute of Technology, Division for
  Geological and Planetary Sciences, MS 150-21, Pasadena, CA 91125,
  USA}
  \altaffiltext{3}{Herschel Science Center, European Space Agency (ESA),
  P.O. Box 78, 28691 Villanueva de la Ca\~nada (Madrid), Spain}
\altaffiltext{4}{Max-Planck Institut f\"ur Extraterrestrische Physik,
  Giessenbachstrasse 1, 85748 Garching, Germany}
\altaffiltext{5}{SRON Netherlands Institute for Space Research,
  P.O. Box 800, 9700 AV Groningen, the Netherlands}
\altaffiltext{6}{University of Toronto, 50 St. George St., Toronto, ON
  M5R 2W9, Canada}
\altaffiltext{7}{Centre for Star and Planet Formation, Natural History
  Museum of Denmark, University of Copenhagen, {\O}ster Voldgade 5-7,
  DK-1350 Copenhagen, Denmark}
\altaffiltext{8}{Laboratoire d'Astrophysique de Grenoble, Universit\'e
  Joseph Fourier, CNRS, UMR 5571, Grenoble, France}

\begin{abstract}

  We present {\sl Spitzer} IRS mid-infrared (5--35 $\mu$m) spectra of
  a complete flux-limited sample ($\geq$ 3 mJy at 8 $\mu$m) of young
  stellar object (YSO) candidates selected on the basis of their
  infrared colors in the Serpens Molecular Cloud. Spectra of 147
  sources are presented and classified. Background stars (with slope
  consistent with a reddened stellar spectrum and silicate features in
  absorption), galaxies (with redshifted PAH features) and a planetary
  nebula (with high ionization lines) amount to 22\% of contamination
  in this sample, leaving 115 true YSOs. Sources with rising spectra
  and ice absorption features, classified as embedded Stage I
  protostars, amount to 18\% of the sample. The remaining 82\% (94) of
  the disk sources are analyzed in terms of spectral energy
  distribution shapes, PAHs and silicate features. The presence,
  strength and shape of these silicate features are used to infer disk
  properties for these systems. About 8\% of the disks have 30/13
  $\mu$m flux ratios consistent with cold disks with inner holes or
  gaps, and 3\% of the disks show PAH emission. Comparison with models
  indicates that dust grains in the surface of these disks have sizes
  of at least a few $\mu$m. The 20 $\mu$m silicate feature is
  sometimes seen in absence of the 10 $\mu$m feature, which may be
  indicative of very small holes in these disks. No significant
  difference is found in the distribution of silicate feature shapes
  and strengths between sources in clusters and in the
  field. Moreover, the results in Serpens are compared with other
  well-studied samples: the c2d IRS sample distributed over 5 clouds
  and a large sample of disks in the Taurus star-forming region. The
  remarkably similar distributions of silicate feature characteristics
  in samples with different environment and median ages -- if
  significant -- imply that the dust population in the disk surface
  results from an equilibrium between dust growth and destructive
  collision processes that are maintained over a few million years for
  any YSO population irrespective of environment.

\end{abstract}


\section{Introduction}
\label{sintro}

Newly formed stars are observed to have infrared (IR) excess due to
their circumstellar disk composed of dust and gas
(\citealt{SS92,HI08}). Most older main-sequence (MS) stars, on the
other hand, have photospheric emission with no excess in the IR. It
is intuitive to conclude that the circumstellar disk evolves with
time, gradually getting rid of the IR excess. One of the main
questions in stellar astrophysics is how this happens.

Observational studies, as well as theoretical simulations, have
demonstrated the interaction between star and disk. The stellar
radiation facilitates disk evolution in terms of photoevaporation
(e.g., \citealt{RI00,RA06,RA08,GH09}) or dust growth and settling
(e.g., \citealt{WE80,ST95,DT97,DD05,JO08}). In the other direction,
mass is accreted from the disk to the star following magnetic field
lines (e.g., \citealt{MU03,WB03,NA04}). A diversity of stellar
temperatures, luminosities and masses among young stars has been known
and studied for decades. Facilitated by new IR and (sub-)millimeter
observations, a great variety of disk shapes, structures and masses is
now being actively studied. The next step is to try to connect stellar
and disk characteristics in order to understand the evolution of these
systems.

The study of a single object, however, is unlikely to provide
unambiguous information regarding the evolutionary stage of the
associated disk. Most studies to date refer to samples of young stars
scattered across the sky, or to sources distributed across large
star-forming clouds like Taurus. In addition to evolutionary stage,
the specific environment in which the stars are formed may influence
the evolution of disks by dynamical and radiative interaction with
other stars or through the initial conditions of the starting cloud,
making it difficult to separate the evolutionary effects (e.g.,
\citealt{RI98,RI00}). For this reason, clusters of stars are very
often used as laboratories for calibrating the evolutionary sequence
(e.g., \citealt{LL95,HA01}). The power of this method, to gain
statistical information on disk composition in coeval samples, was
found to be very successful for loose associations of older, pre-main
sequence stars such as the 8 Myr old $\eta$ Cha \citep{BO06} and the
10 Myr old TW Hydrae association \citep{UC04}. Identifying clusters of
even younger disk populations is a natural step towards the completion
of the empirical calibration of the evolution of disks surrounding
young low-mass stars. This paper analyzes the inner disk properties of
a flux-limited, complete unbiased sample of young stars with IR excess
in the Serpens Molecular Cloud (d = 259 $\pm$ 37 pc, \citealt{ST96})
which has a mean age $\sim 5$ Myr, \citep{OL09} with an YSO population
in clusters and also in isolation. It has been recently argued
(L. Loinard, private communication) that the distance to Serpens could
be considerably higher than previously calculated. This would imply a
rather younger median age for this cloud.

The Spitzer Legacy Program ``From Molecular Cores to Planet-Forming
Disks'' (c2d) has uncovered hundreds of objects with IR excess in five
star-forming clouds (Cha II, Lupus, Ophiuchus, Perseus and Serpens),
and allowed statistical studies within a given cloud \citep{EV09}. The
c2d study of Serpens with IRAC (3.6, 4.5, 5.8 and 8.0 $\mu$m,
\citealt{FA04}) and MIPS (24 and 70 $\mu$m, \citealt{RI04}) data has
revealed a rich population of mostly previously unknown young stellar
objects (YSOs) associated with IR excess, yielding a diversity of disk
SEDs (\citealt{HA06,HB07,HA07}). Because of the compact area in
Serpens mapped by Spitzer (0.89 deg$^2$), this impressive diversity of
disks presents itself as an excellent laboratory for studies of early
stellar evolution and planet formation. Indeed, the Serpens core
(Cluster A, located in the northeastern part of the area studied by
c2d) has been well studied in this sense (e.g.,
\citealt{ZA88,EC92,TS98,KA04,EI05,WI07,WI09}), whereas only some of
the objects in Group C (formerly known as Cluster C) were studied with
ISOCAM data \citep{DJ06}.

Because of its wavelength coverage, sensitivity and mapping
capabilities, the Spitzer Space Telescope has offered an opportunity
to study many of these systems (star$+$disk) in unprecedented
detail. Spitzer's photometry in the mid-IR, where the radiation
reprocessed by the dust is dominant, gives information on the shape of
the disks and, indirectly, its evolutionary stage (assuming an
evolution from flared to flat disks). Follow-up mid-IR spectroscopic
observations with the InfraRed Spectrograph (IRS, 5 -- 38, $\mu$m
\citealt{HO04}) on-board Spitzer probe the physical and chemical
processes affecting the hot dust in the surface layers of the inner
regions of the disk. The shapes and strengths of the silicate features
provide information on dust grain size distribution and structure
(e.g., \citealt{VB03,PY03,BO08,KE06,KE07,GE06,WA09,OF09}). These, in
turn, reflect dynamical processes such as radial and vertical mixing,
and physical processes such as annealing. A smooth strong Si--O
stretching mode feature centered at 9.8 $\mu$m is indicative of small
amorphous silicates (like those found in the ISM) while a structured
weaker and broader feature reveals bigger grains or the presence of
crystalline silicates. Polycyclic aromatic hydrocarbon (PAH) features
are a probe of the UV radiation incident on the disk, whereas their
abundance plays a crucial role in models of disk heating and chemistry
(e.g., \citealt{GE06,DU07,VI07}). The shape and slope of the
mid-infrared excess provides information on the flaring geometry of
the disks \citep{DD04}, while ice bands may form for highly inclined
sources (edge-on) where the light from the central object passes
through the dusty material in the outer parts of the disk
\citep{PO05}. Thus, the wavelength range probed by the IRS spectra
enables analysis of the geometry of individual disks. It also probes
the temperature and dust size distributions as well as crystallinity
of dust in the disk surface at radii of 0.1 -- few AU. Statistical
results from a number of sources help the understanding of the
progression of disk clearing and possibly planet formation.

Our group has been conducting multi-wavelength observing campaigns of
Serpens. Optical and near-IR wavelength data, where the stellar
radiation dominates, are being used to characterize the central
sources of these systems \citep{OL09}. Effective temperatures,
luminosities, extinctions, mass accretion rates, as well as relative
ages and masses are being determined. In this paper, we present a
complete flux-limited set of Spitzer IRS spectra for this previously
unexplored young stellar population in Serpens. We analyze these
spectra in terms of common and individual characteristics and compare
the results to those of Taurus, one of the best studied molecular
clouds to date and dominated by isolated star formation. A subsequent
paper will deal with the full SED fitting for the disk sources in
Serpens and their detailed analyzes. Our ultimate goal is to
statistically trace the evolution of young low-mass stars by means of
the spectroscopic signatures of disk evolution, discussed above.

Section \ref{sdata} describes our Spitzer IRS data. In \S~\ref{sres}
we divide our sample into categories based on their IRS spectra: the
background contaminants are presented in \S~\ref{sbg} separated
according to the nature of the objects; embedded sources are presented
in \S~\ref{semb}, and disk sources in \S~\ref{sdisks} (with emphasis
on PAH and silicate emission). In \S~\ref{sdis} we discuss disk
properties in relation to environment and to another cloud, Taurus,
for comparison. In \S~\ref{scon} we present our conclusions.

\section{Spitzer IRS Data}
\label{sdata}

\subsection{Sample Selection}
\label{ssample}

\citet{HA07} describe the selection criteria used by the c2d team to
identify YSO candidates based on the Spitzer IRAC and MIPS data,
band-merged with the 2MASS catalog. They used a number of color-color
and color-magnitude diagrams to separate YSOs (both embedded YSOs and
young stars with disks) from other types of sources, such as background
galaxies and stars. In this manner, a set of 235 YSOs was identified
in Serpens, in both clusters and in isolation, as described in
\S~\ref{scf}. These criteria, however, are not fail-proof. \citet{HA07}
treated this problem very carefully in a statistical sense, but the
locus region for YSOs and, for instance, background galaxies overlap
somewhat in color-color diagrams with Spitzer photometry. Therefore,
some contamination may well still be present in the sample which was
impossible to disentangle from photometry alone.

\begin{figure}[ht!]
\begin{center}
\includegraphics[width=0.45\textwidth]{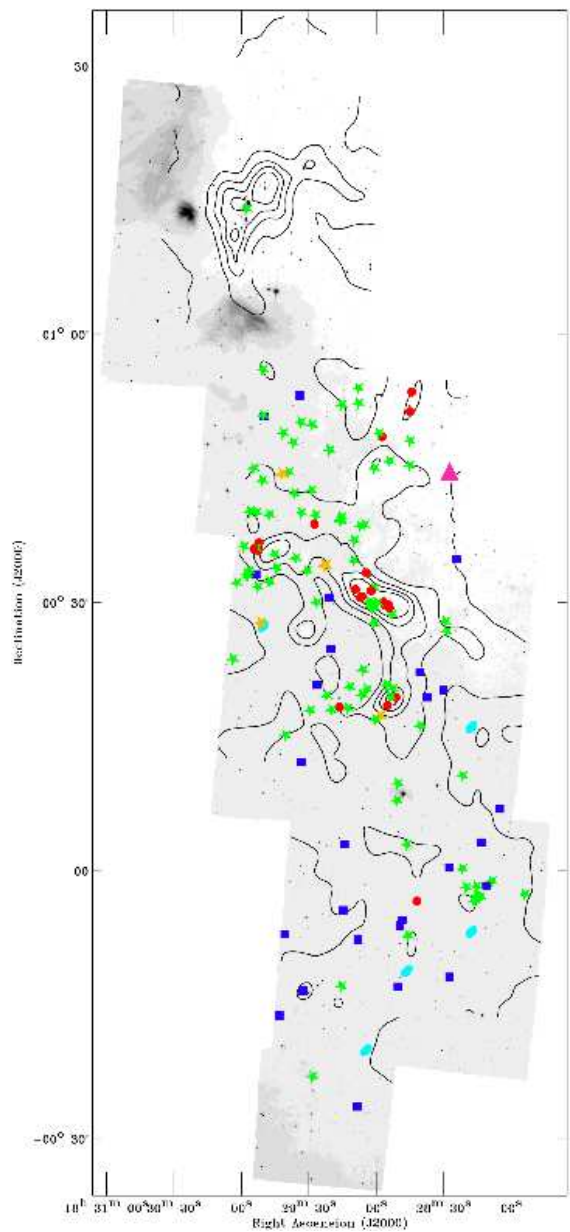}
\end{center}
\caption{\label{objirac} Observed objects over-plotted on the c2d
  IRAC4 (8.0 $\mu$m) map of Serpens. The contours (5, 10, 15, 20 and
  25 mag) of visual extinction are derived from the c2d extinction
  maps \citep{EV07}. Blue squares are background stars, cyan ellipses
  are redshifted galaxies, red circles are embedded sources, and stars
  represent disks (yellow disks with PAH emission and green other disk
  sources). The pink triangle is the PN candidate, see \S~\ref{sbgpn}
  for details.}
\end{figure}

An additional criterion was applied to this sample in order to
guarantee IRS spectra with sufficient quality to allow comparative
studies of solid-state features, namely S/N $\geq$ 30 on the
continuum. A lower limit flux cutoff of 3 mJy at 8$\mu$m was
imposed. This flux limit ensures coverage down to the brown dwarf
limit ($L \sim$ 0.01 L$_\odot$) and leads to a final sample of 147
objects, the same as in \citet{OL09}. This is a complete flux-limited
sample of IR excess sources in the c2d mapped area, except in the
Serpens Core. It is also important to note that this sample, by
definition, does not include young stars without IR excess (Class
III). Figure \ref{objirac} shows the distribution of our observed
objects in the Serpens Molecular Cloud. Table \ref{tab1} lists the
sample together with their 2--24 $\mu$m spectral slope.

\subsection{Observations and Data Reduction}
\label{sobs}

Out of the 147 objects in this sample, 22 were observed between
September 2004 and June 2006, as part of the c2d IRS 2nd Look
campaign. The reduced high resolution spectra (R = $\lambda/ \Delta
\lambda$ = 600; Short High [SH], 9.9 -- 19.6 $\mu$m and Long High
       [LH], 18.7 -- 37.2 $\mu$m), complemented with Short Low spectra
       (SL, R = 50 -- 100; 5.2 -- 14.5 $\mu$m) are public to the
       community through the c2d delivery website \citep{EV07}. The
       other 125 targets were observed as part of a Spitzer Space
       Telescope's cycle 3 program (GO3 \#30223, PI: Pontoppidan) in
       the low resolution module (SL and Long Low [LL], 14.0 -- 38.0
       $\mu$m) in 35 AORs, between October 2006 and April 2007.

All objects were observed in IRS staring mode and extracted from the
SSC pipeline (version S12.0.2 for the c2d IRS 2nd look program, and
version S15.3.0 for the GO3 program) basic calibration data (BCD)
using the c2d reduction pipeline \citep{LH06}. This process includes
bad-pixel correction, optimal PSF aperture extraction, defringing and
order matching. Technical information about each object can be found
in Table \ref{tab1}.

\section{Results}
\label{sres}

The large number of objects studied here allows statistical studies of
each stage of disk evolution in a single star-forming region. By
simply looking at the spectra, a diversity of spectral shapes can be
noted immediately. This diversity of objects can be divided into
several categories, each of which is thought to be related to a
different evolutionary stage of the star$+$disk system, described
below.

\subsection{Background Sources}
\label{sbg}

Due to the low galactic latitude of Serpens ($l$=030.4733,
$b$=+05.1018) the contamination by galactic background sources is
expected to be higher for this region than for other studied
star-forming regions (e.g., Taurus, Lupus, Chamaeleon). With the help
of the IRS spectra, background contaminants can readily be identified,
even when seen through a large column density of molecular cloud
material. These are presented and discussed in Appendix \ref{abg}.

To further study the YSOs, the 32 contaminants are removed from the
sample, which then consists of 115 objects for further analysis.

\subsection{Embedded Sources}
\label{semb}

A newly formed protostar, still embedded in an envelope of gas and
dust, will show an excess in the mid-IR that rises with wavelength. In
this cold environment, a significant fraction of the molecules are
found in ices. Many of these ices are observed as absorption features
in the wavelength range of the IRS instrument (e.g., \citealt{AB04}).

\begin{figure}[th!]
\begin{center}
\includegraphics[width=\textwidth]{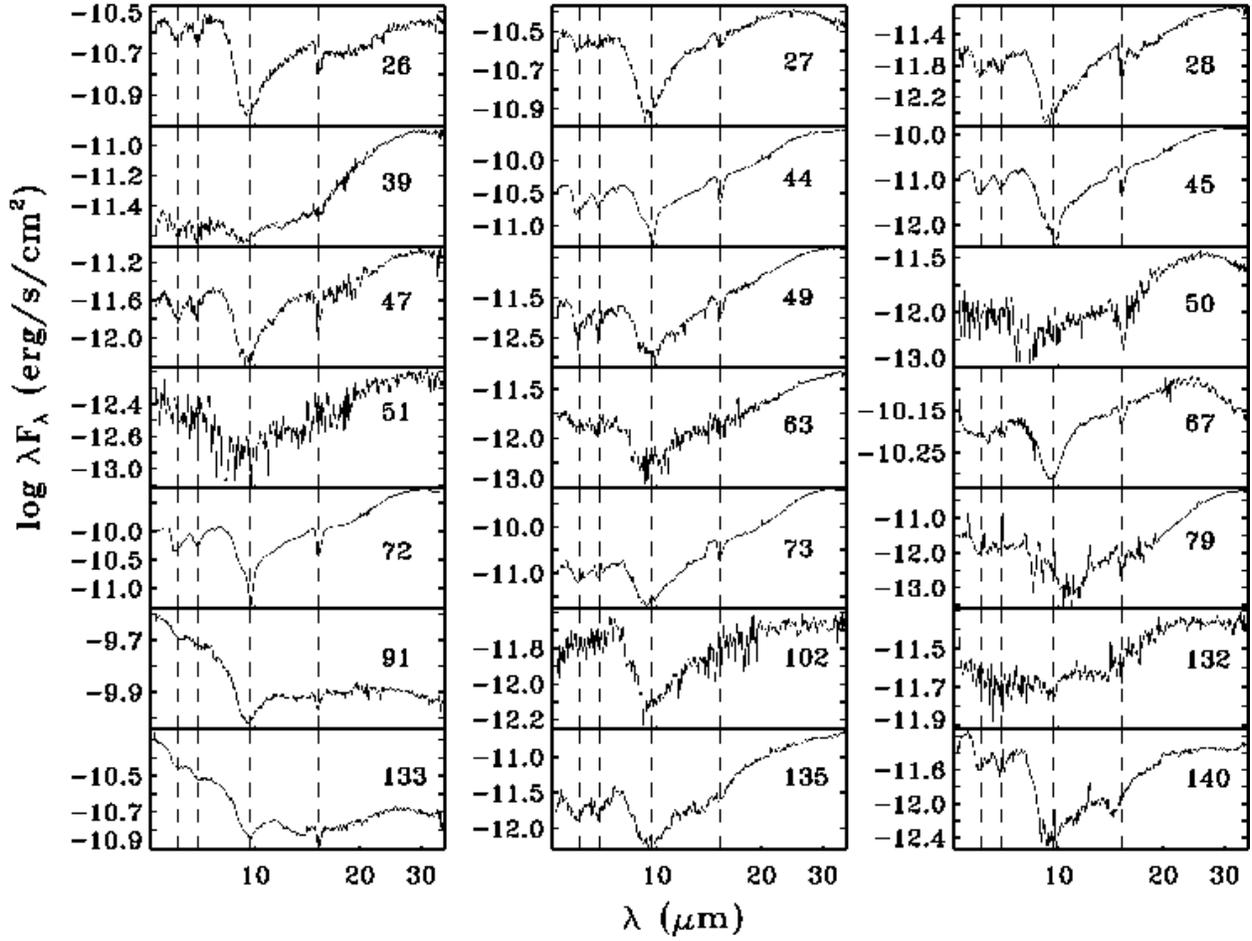}
\end{center}
\caption{\label{irsemb} IRS spectra of the embedded sources in Serpens
  with rising mid-IR spectra. The prominent ice absorption features
  (at 6.0, 6.85, 7.7, 9.0 and 15.2 $\mu$m) are marked, together with
  the silicate absorption feature at 9.7 $\mu$m. The object \# is
  indicated at the top right of each panel.}
\end{figure}

Of our sample, 21 sources have spectra consistent with embedded
sources. These spectra are shown in Figure \ref{irsemb}. The most
prominent ice feature, CO$_2$ at 15.2 $\mu$m, can readily be seen in
all spectra. Weaker features, such as H$_2$O at 6.0 and 6.85 $\mu$m
and CH$_3$OH at 9.7 $\mu$m, need high S/N. Those features are detected
in objects \#26, 27, 28, 44, 45, 47, 49, 67, 72, 73, 135 and 140. As
can be seen in Figure \ref{objirac}, the embedded objects seem to be
concentrated in high extinction regions ($A_V \sim 10-15$ mag).

The embedded objects observed as part of the c2d IRS 2nd Look Program
have been studied in detail in a series of papers on ices around YSOs
(H$_2$O in \citealt{AB08}; CO$_2$ in \citealt{PO08}; CH$_4$ at 7.7
$\mu$m in \citealt{OB08}; and NH$_3$ at 9.0 $\mu$m and CH$_3$OH in
\citealt{SB09}) and are not discussed further in this paper.

\subsection{Disk Sources}
\label{sdisks}

Young stellar objects with IR excess that have a flat or negative
slope in the near- to mid-IR are identified as disk sources. Following
\citealt{GR94} and \citealt{EV09}, flat sources have $-0.3 \leq
\alpha_{2\mu{\rm m}-24\mu{\rm m}} \leq 0.3$, Class II sources have
$-1.6 \leq \alpha_{2\mu{\rm m}-24\mu{\rm m}} \leq -0.3$, and Class III
sources have $\alpha_{2\mu{\rm m}-24\mu{\rm m}} \le -1.6$. Many of
these sources also have silicate emission bands and a few show PAH
features. In the following, these different features are analyzed and
discussed in detail for the total of 94 disk sources.

\subsubsection{Disk Geometry: $F_{30}/F_{13}$ and Cold Disks}
\label{scd}

\begin{figure}[h!]
\begin{center}
\includegraphics[width=0.5\textwidth]{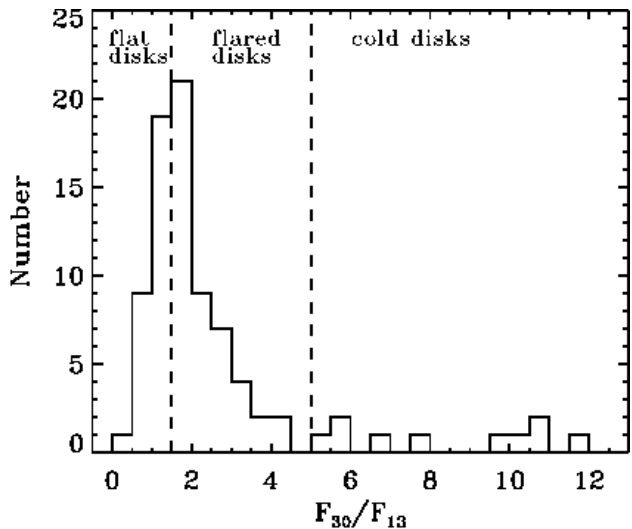}
\end{center}
\caption{\label{irsshape} Distribution of the flux ratio between 30
  and 13 $\mu$m, used as a proxy for disk geometry.}
\end{figure}

Disk geometry is inferred from the flux ratio between 30 and 13 $\mu$m
($F_{30}/F_{13}$). High ratios ($F_{30}/F_{13} \gtrsim 15$) yield
edge-on disks, intermediate values ($5 \lesssim F_{30}/F_{13} \lesssim
15$ and $1.5 \lesssim F_{30}/F_{13} \lesssim 5$) identify cold disks
and flared disks with considerable excess in the IR, respectively, and
low ratios ($F_{30}/F_{13} \lesssim 1.5$) select flat, settled disks
with little IR excess \citep{JB07,ME09}. Figure \ref{irsshape} shows
the distribution of $F_{30}/F_{13}$ for the disk population in
Serpens.

Cold disks, sometimes referred to as transitional disks, as studied in
\citet{JB07} and \citet{ME09}, present a peculiar spectral energy
distribution (SED), with inner gaps or holes producing a region with
no IR excess in the near- to mid-IR but a substantial excess at mid-
to far-IR wavelengths. Binaries \citep{IK08}, planet formation
\citep{QU04,VR06} and photoevaporation \citep{CL01,RA06} are some of
the suggested mechanisms for this removal of warm dust.

\begin{figure}[h!]
\begin{center}
\includegraphics[width=\textwidth]{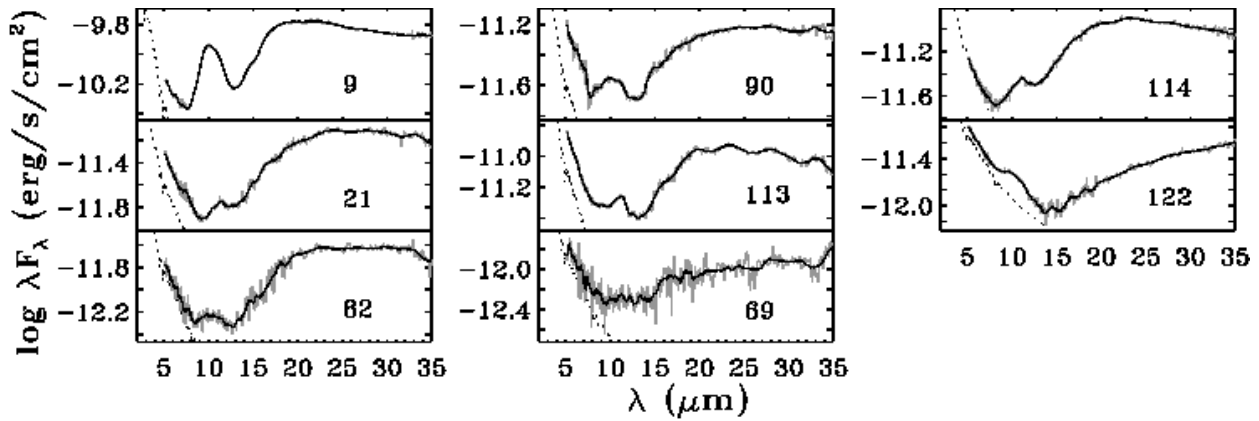}
\end{center}
\caption{\label{irscd} IRS spectra of the cold disks in
  Serpens. Object \# is indicated in each panel. In grey are the
  original spectra, while in black a binned version is
  overplotted. The dotted line represents the stellar photosphere, for
  comparison. }
\end{figure}

Following \citet{JB07}, $F_{30}/F_{13}$ is used to differentiate cold
disks from the general disk population. Those wavelengths were chosen
to avoid the silicate features and to probe the spectra steeply rising
at mid- to far-IR following a deficit at shorter wavelengths,
characteristic of cold disks. Eight objects (\# 9, 21, 69, 82, 90,
113, 114 and 122) in this sample are classified as cold disks based on
$5 \lesssim F_{30}/F_{13} \lesssim 15$, corresponding to 8.5\% of the
disk population. These spectra are shown in Figure \ref{irscd}. For
clarity, the original spectra are shown in grey, while in black a
binned version is overplotted. The binning was done using a
Savitzky-Golay \citep{SG64} filter of order 0 and degree 3, for
illustrative purposes only. This is the most unbiased survey of cold
disks to date, based entirely on spectroscopy in the critical mid-IR
range, and should thus give a representative fraction of such
disks. These objects are also part of a larger cold disk sample,
studied by \citet{ME09}, where the authors have selected cold disks
from photometric criteria in all five c2d clouds.

\subsubsection{PAH Emission}
\label{spah}

Polycyclic aromatic hydrocarbons (PAH) are large molecules whose
emission is excited by UV radiation out to large distances from the
star. Thus, these features are good diagnostics of the amount of
stellar UV intercepted by the disk surface.

Almost half of the sample of Herbig Ae/Be stars (young stars of
intermediate masses) studied by \citet{ME01} and \citet{AA04} show PAH
emission. \citet{GE06}, on the other hand, show that PAH emission is
not as common among low-mass YSOs as it is for more massive stars. The
authors argue that the lack of features is not only due to their
weaker UV radiation fields, but also to a lower PAH abundance compared
with the general ISM.

\begin{figure}[h!]
\begin{center}
\includegraphics[width=0.5\textwidth]{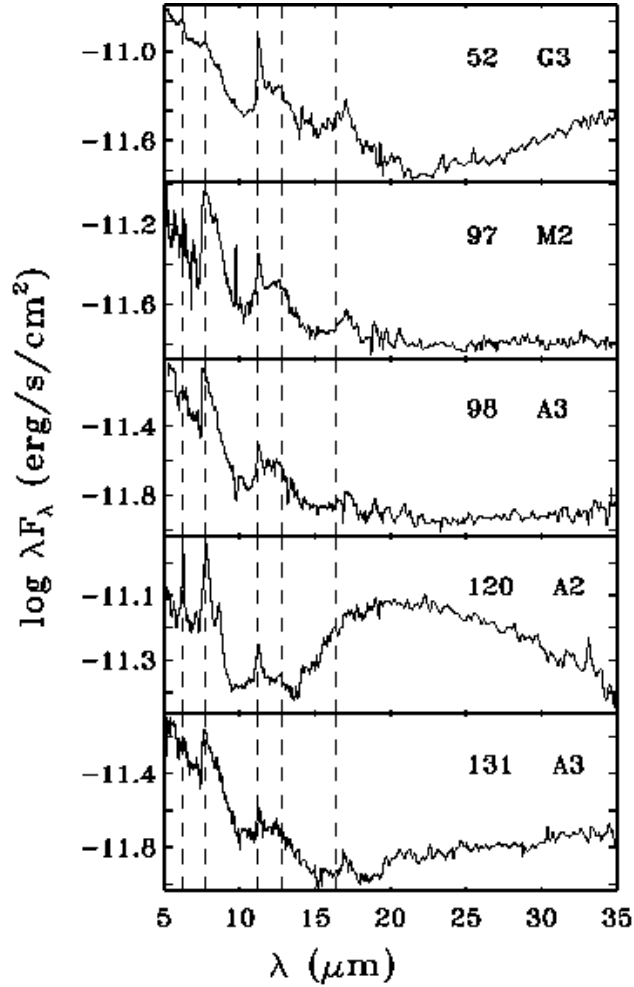}
\end{center}
\caption{\label{irspah} IRS spectra of the disk sources which show
  clear PAH features. The dashed lines indicate the positions of the
  most prominent bands (at 6.2, 7.7, 8.6, 11.2, 12.8 and 16.4
  $\mu$m). The object \# and its spectral type (from \citealt{OL09})
  are indicated at the top right of each panel. Note that the ratios
  of different line strengths differ between sources. }
\end{figure}

In the current sample, 5 sources have clear PAH emission, presented in
Figure \ref{irspah}. Although a great variation in shapes and
strengths is seen, all sources show the features at 6.2, 7.7, 8.6,
11.2 and 12.8 $\mu$m. Objects \#52, 97 and 131 also show a feature
that could be the PAH band at 16.4 $\mu$m, although the feature does
not match perfectly in position.

The central stars in these 5 sources were studied with optical
spectroscopy in \citet{OL09}. Spectral types and luminosities were
derived, yielding ages and masses for each object. Objects \#52, 98,
120 and 131 have masses between 1.9 and 2.7 $M_\odot$ (spectral type
G3 to A2), while object \#97 has lower mass, of around 0.45 $M_\odot$
(M2). Objects \#97 and 98 are spatially very close to each other and
have very similar PAH features. IRAC Band 4 (8 $\mu$m) images show
that there is extended nebulosity surrounding both positions so that
the PAHs are likely associated with this nebula rather than the disks.

Discarding these 2 sources, this sample indicates that PAHs are
present in 3.2\% of the disk population studied, which is below the
11--14\% of PAH detection rate found by \citet{GE07}. This discrepancy
could be due to the two samples being differently selected, with the
current sample having a larger fraction of late-type stars. The
percentage of PAHs present in disks in Serpens is, however, very
comparable with that in Taurus (4\% of the disk population,
\S~\ref{stau}).

Interestingly, none of the cold disks show PAHs, different from
expected if grain settling enhances PAH features \citep{DU07} and
observed for other cold disk samples studied which do show PAHs
\citep{JB07,ME09}. In addition, the objects with PAH emission have
standard $F_{30}/F_{13}$ flux ratios, falling in the `flat' and
`flared' regions of Figure \ref{irsshape}.

Only object \#120 shows a silicate feature (at 20 $\mu$m, as discussed
in \S~\ref{ssi}), in addition to PAHs. This alludes to an
anti-correlation between PAHs and silicates that could be simply
explained in terms of contrast, i.e., it is hard to detect PAH
features when a strong 10 $\mu$m silicate feature is present
\citep{GE06}. In this sense, PAHs should be more easily detected in
disks without silicates, since the falling spectrum provides an
increase in the feature to continuum ratio improving
contrast. 

\subsubsection{Silicate Emission}
\label{ssi}

\begin{figure}[h!]
\begin{center}
\includegraphics[width=\textwidth]{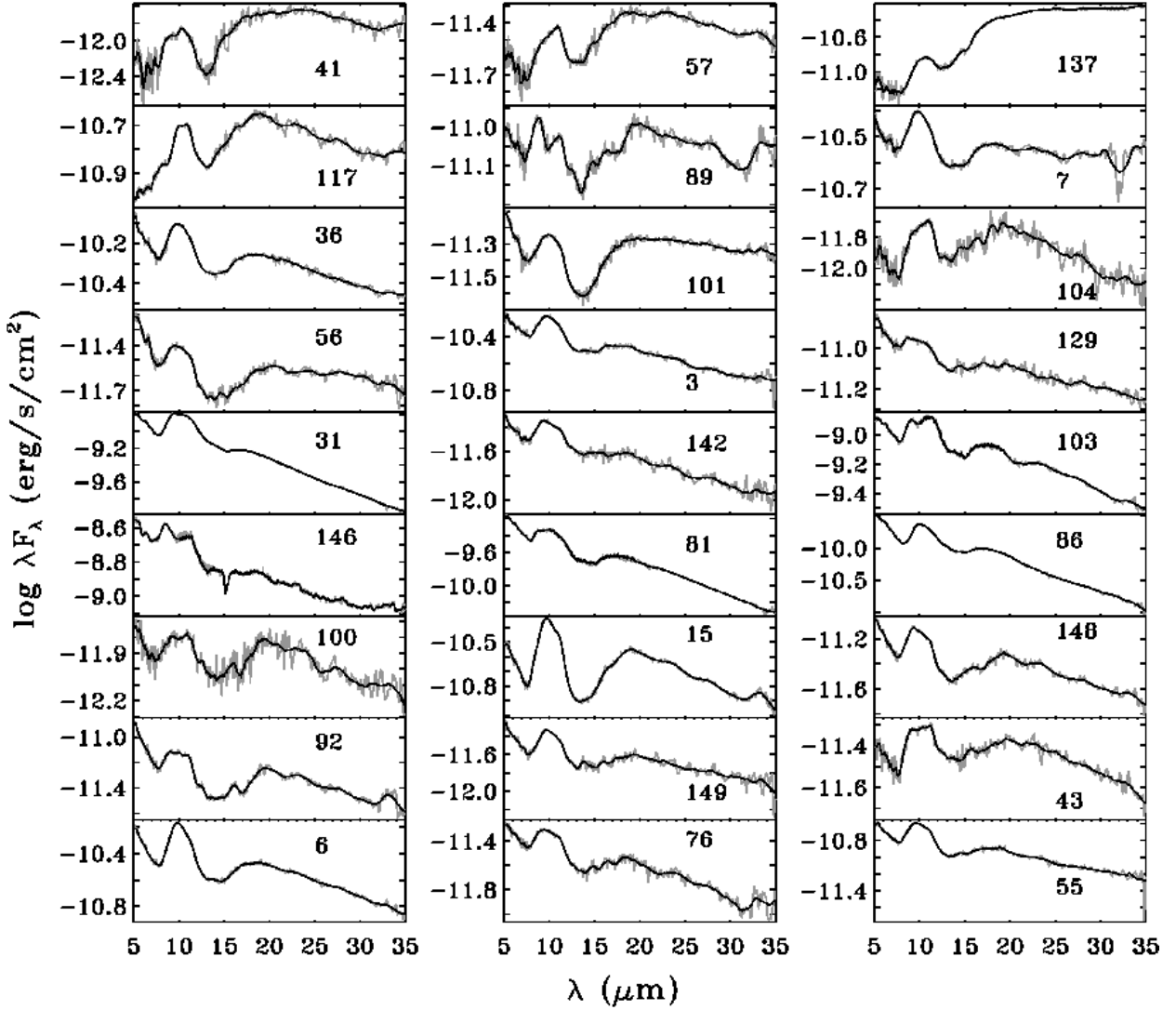}
\end{center}
\caption{\label{irsdisk1} IRS spectra of the disk sources in
  Serpens. Object \# is indicated in each panel. In grey are the
  original spectra, while in black a binned version (using a
  Savitzky-Golay \citep{SG64} filter of order 0 and degree 3) is
  overplotted. }
\end{figure}

\begin{figure}[h!]
\begin{center}
\includegraphics[width=\textwidth]{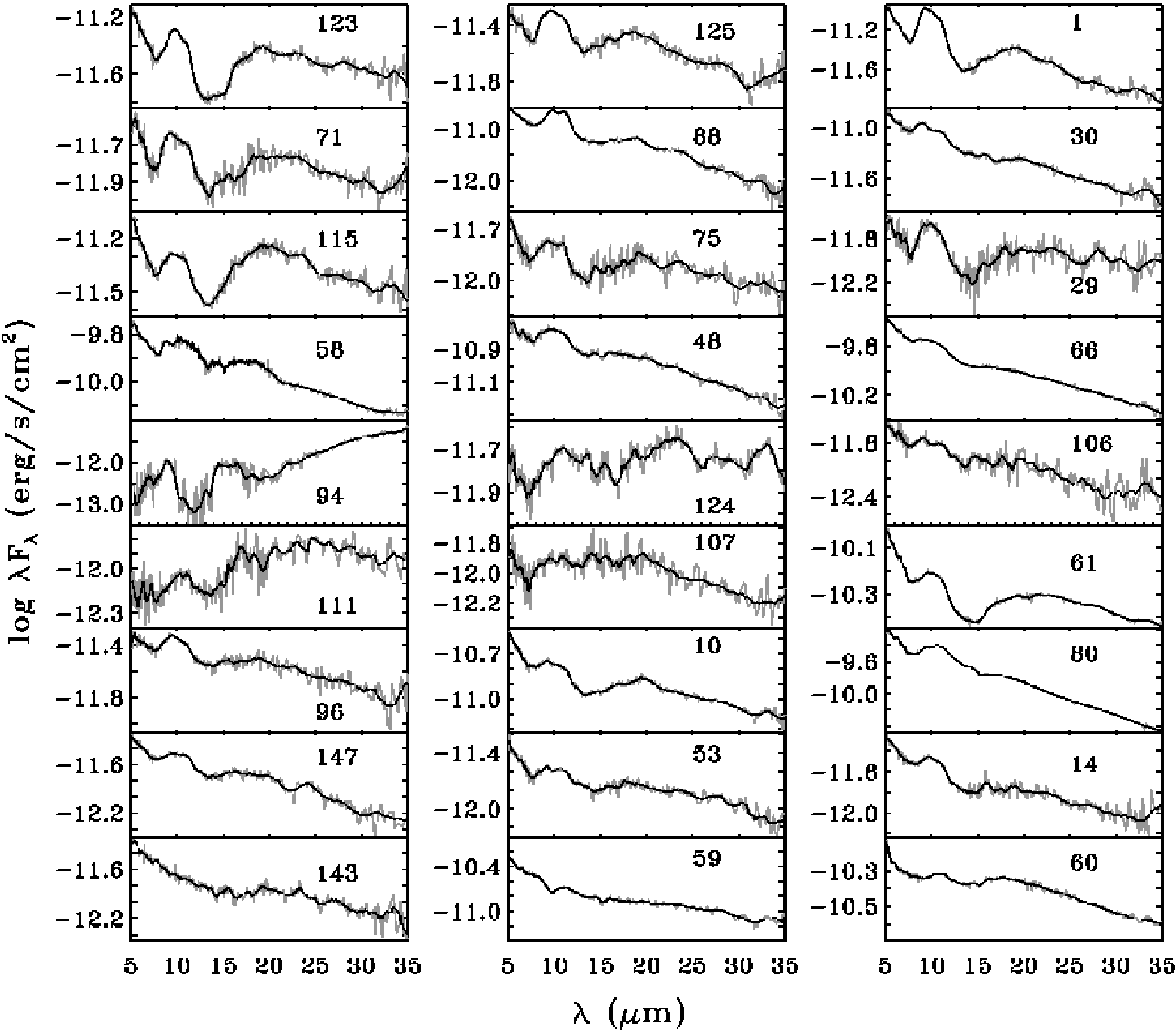}
\end{center}
\caption{\label{irsdisk2} IRS spectra of the disk sources in Serpens, contd. }
\end{figure}

\begin{figure}[h!]
\begin{center}
\includegraphics[width=\textwidth]{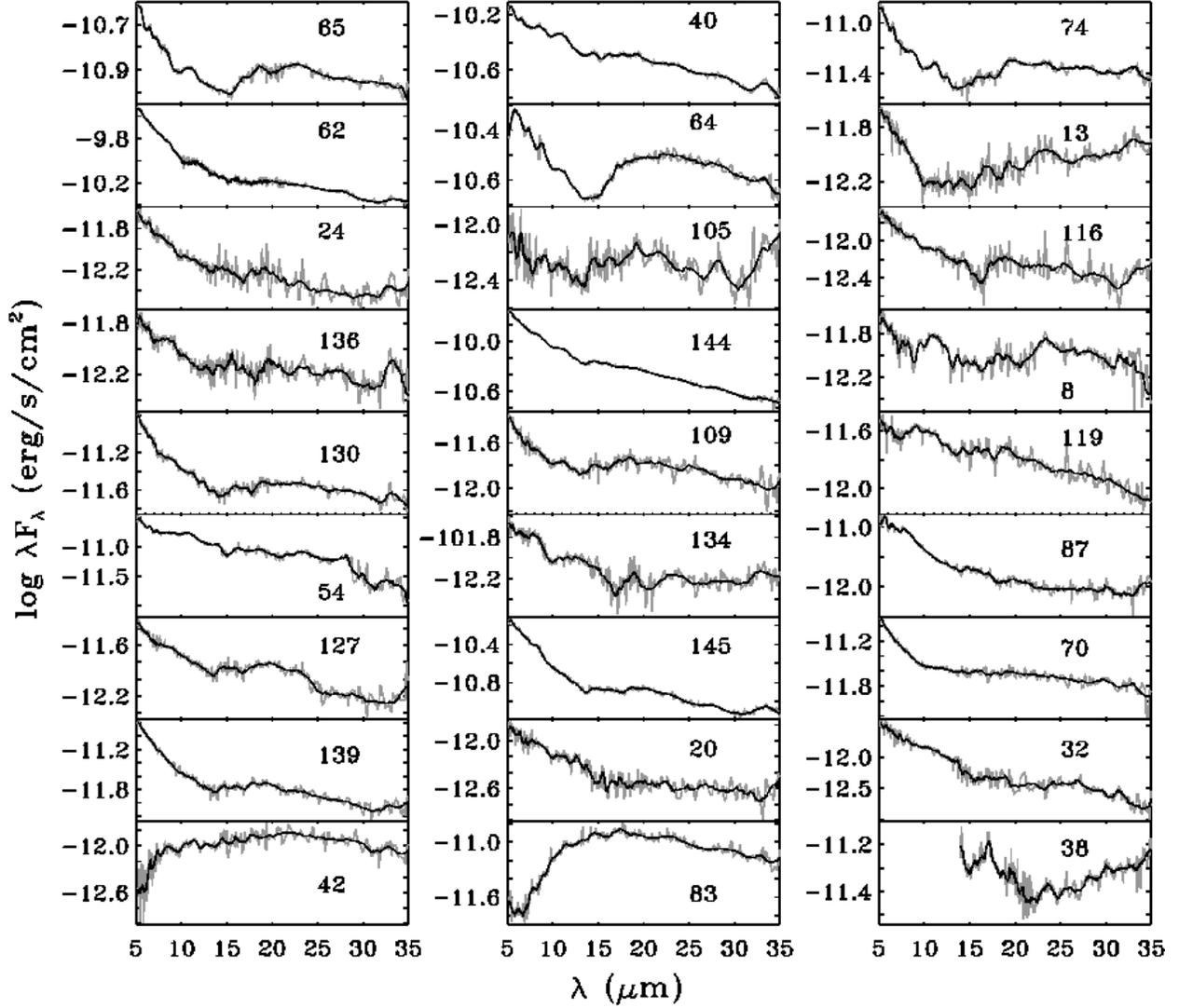}
\end{center}
\caption{\label{irsdisk3} IRS spectra of the disk sources in Serpens, contd. }
\end{figure}

Silicate emission has been observed around numerous circumstellar
disks (e.g., \citealt{VB03,PY03,KE06,FU06,BO08,OF09}). Some of this
material has different characteristics in disks compared to primitive
interstellar material. The strength and the shape of silicate features
(at 10 and 20 $\mu$m) contain important information on the dust sizes
in the surface layer of the disk. The 89 disk sources in this sample
were inspected for their silicate characteristics. Their spectra can
be seen in Figures \ref{irsdisk1} to \ref{irsdisk3} (completed with
Figures \ref{irscd} and \ref{irspah}), arranged in order of decreasing
strength of the 10 $\mu$m feature and decreasing $\alpha_{2-24\mu{\rm
    m}}$ spectral slope. No extinction correction has been applied to
these spectra (see \S~\ref{ssi_ext}).

Five of these sources (5\% of the disk population) have spectra that
are featureless, i.e., they have neither the 10 nor the 20 $\mu$m
silicate features in emission, although they present IR excess
(objects \# 38, 42, 59, 83, and 87). It is found that when the 10
$\mu$m feature appears, the 20 $\mu$m band is also present (by visual
inspection of the continuum-subtracted spectra, see
\S~\ref{ssi_ssr}). The contrary is not true, as objects \#13, 20, 24,
32, 62, 64, 69, 70, 109, 116, 127, 139, 143 and 145 have an observed
20 $\mu$m feature but no apparent 10 $\mu$m feature (also reported by
\citet{KE06} for T Cha, SR21 and HD 135344B). Also, object \#120 shows
the 20 $\mu$m band and no 10 $\mu$m feature, as this region is
dominated by PAH features in this source (see \S~\ref{spah}). When
present, the 10 $\mu$m feature ($S^{10\mu{\rm m}}_{{\rm peak}}$) is
usually stronger than the 20 $\mu$m ($S^{20\mu{\rm m}}_{{\rm peak}}$)
one as is common in disk sources.

Crystalline silicate features (as discussed by
e.g. \citealt{BO01,BO08,AA04,VB04,AP05,OF09,ST09}) are seen in several
sources. A detailed analysis in terms of statistics and crystalline
fraction of these features is non-trivial and beyond the scope of this
paper, but will be subject of a future publication.

\subsubsection{The Silicate Strength-Shape Relation}
\label{ssi_ssr}

As first shown for Herbig Ae stars \citep{VB03}, there is a dependence
between grain size and silicate shape and strength for both 10 and 20
$\mu$m features: as the sizes of the grains grow, the silicate
features appear more and more flattened and shift to longer
wavelengths.

A statistical analysis of these two features can be done using the
following quantities: $S^{10\mu{\rm m}}_{{\rm peak}}$ is a measure of
the strength of the feature centered at 10 $\mu$m and
$S_{11.3}/S_{9.8}$ is a proxy for the shape of the feature (the
smaller this value, the more peaked the feature is); $S^{20\mu{\rm
    m}}_{{\rm peak}}$ measures the strength of the 20 $\mu$m feature,
while $S_{23.8}/S_{19}$ is a proxy for this feature shape. Following
\citet{KE06}, $S_{\lambda}$ is defined as:

\begin{equation}
\label{es}
S_\lambda = 1 + \frac{1}{2\delta}\int_{\lambda-\delta}^{\lambda+\delta}\frac{F_\lambda - C_\lambda}{C_\lambda} ~d \lambda.
\end{equation}

In this paper, we use $\delta = 0.1\,\mu$m.

\begin{figure}[h!]
\begin{center}
\includegraphics[width=\textwidth]{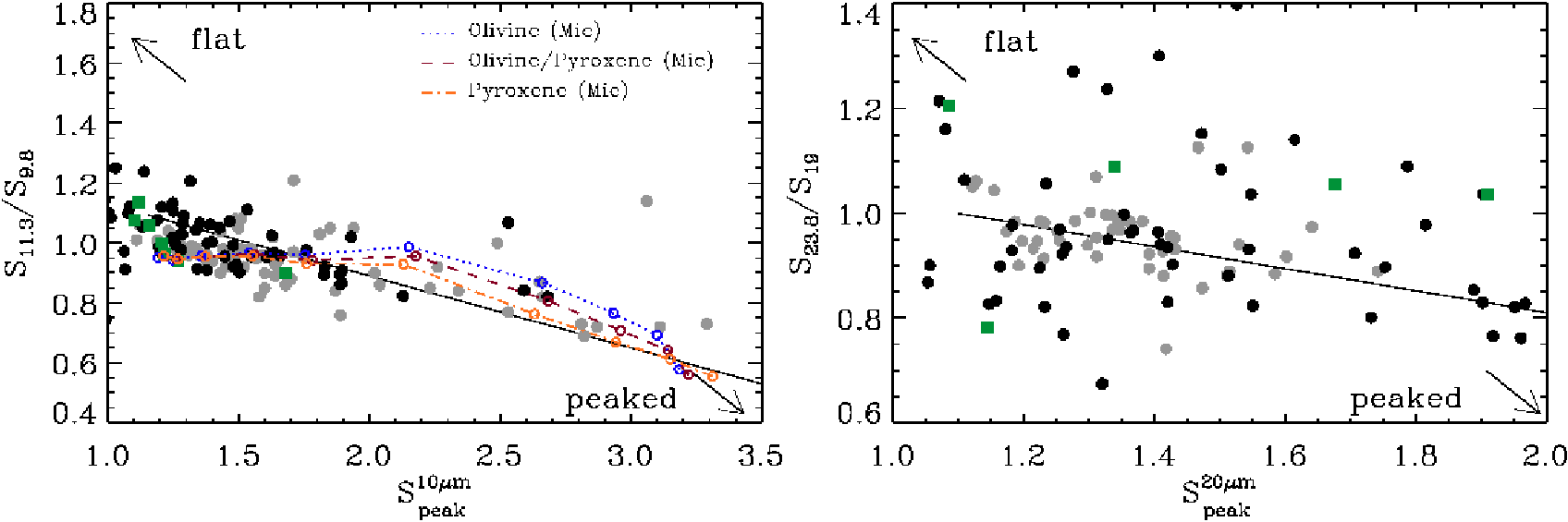}
\end{center}
\caption{\label{si} In the left panel, black dots show the ratio of
  peaks at 11.3 to 9.8 $\mu$m ($S_{11.3}/S_{9.8}$) plotted against the
  peak at 10 $\mu$m ($S^{10\mu{\rm m}}_{{\rm peak}}$). In the right
  panel, the ratio of peaks at 23.8 to 19 $\mu$m ($S_{23.8}/S_{19}$)
  is plotted against the peak at 20 $\mu$m ($S^{20\mu{\rm m}}_{{\rm
      peak}}$). The best fit found by \citet{KE06} for the c2d IRS
  first look sample of YSOs (grey dots) has been indicated for
  comparison (solid black line). Green squares are the cold disks in
  this sample (see \S~\ref{scd}). Colored curves are derived from
  theoretical opacities for different mixtures by \citet{OF09}. The
  open circles correspond to different grain sizes, from left to right
  6.25, 5.2, 4.3, 3.25, 2.7, 2.0, 1.5, 1.25, 1.0 and 0.1
  $\mu$m. Typical uncertainties for peak strength are $\sim$0.1, while
  typical uncertainties for the peak ratios are $\sim$0.12. }
\end{figure}

\citet{KE06} found a global trend (best fit to observations) for both
features (Figure 9 in their paper and \ref{si} in this one),
indicating that the feature shape and strength are physically related,
as expected if the silicate features are indeed tracers of grain
size. The relationship is much tighter for the 10 $\mu$m feature, as
also seen by the great spread in the correlation for the 20 $\mu$m
feature.

Following their procedure, we first subtracted the continuum of each
spectrum by fitting a second order polynomial to the following
regions: blue-ward of the 10 $\mu$m feature, between the 10 and the 20
$\mu$m features, and red-ward of the 20 $\mu$m feature (6.8--7.5,
12.5--13.5 and 30--36 $\mu$m). The peak fluxes at 9.8, 10, 11.3, 19,
20 and 23.8 $\mu$m were then determined (see Table \ref{tsi}). Figure
\ref{si} shows the results for our sample of disks, excluding those
with PAH emission. The best fit found by \citet{KE06} for the c2d IRS
first look sample of YSOs has been indicated for comparison (solid
line). The agreement of the Serpens data with the best fit of
\citet{KE06} is excellent, and more evident for the 10 $\mu$m than for
the 20 $\mu$m feature due to the larger scatter in the latter group.

\citet{OF09} generated synthetic 10 $\mu$m features calculated for
different grain sizes and compositions. Their models are generated for
amorphous silicates of olivine and pyroxene stoichiometry and a 50:50
mixture, with grain size varying between 0.1 and $\sim$6 $\mu$m. Those
models are overplotted in the left panel of Figure \ref{si}, with open
symbols corresponding to different grain sizes. The comparison of the
data presented here with these models shows that the majority of our
sample lies in a region consistent with an opacity dominated by grains
with sizes larger than 2.0 $\mu$m, having no features consistent with
grain sizes smaller than 1.5 $\mu$m. A considerable fraction of the
sample is consistent with grains as large as 6 $\mu$m (the largest
grain size modeled). The precise sizes depend on composition and
treatment of the opacities, but sizes of a few $\mu$m are clearly
implicated. It is important to note that crystallinity also affects
the shape of silicate features. However, as shown by \citet{AP05} and
\citet{OF09}, the effect of crystallinity is orthogonal to that of
grain sizes in the strength versus shape plot (Figure \ref{si}). Grain
sizes are found to be the dominant parameter changing the shape of the
10 $\mu$m silicate feature, whereas crystallinity introduces scatter
(Figure 13 of \citealt{OF09}).

All cold disks present silicate features in emission. According to the
models of grain sizes (left panel of Figure \ref{si}), the cold disks
in this sample have grains bigger than 2.7 $\mu$m, with the bulk of
the sample presenting grains as big as 6.0 $\mu$m, according the these
models.

\subsubsection{Effects of Extinction}
\label{ssi_ext}

The amount of foreground extinction is, for some sources, substantial
enough to affect the appearance of silicate emission features in the
mid-infrared. Because the extinction law has strong resonances from
silicates, the strength and shape of any silicate emission feature may
be significantly affected if a large column of cloud material is
present in front of a given source. Here, the potential effects of
extinction on statistical results, such as those presented in Figure
\ref{si}, are discussed.

The model presented by \cite{WD01} for an absolute to selective
extinction of $R_V=5.5$ has been found to be a reasonable match to the
observed dark cloud extinction law \citep{CH09,MC09,CM09}, and it is
assumed that this law holds for Serpens. One caveat is the lack of
resonances due to ices in this model, but given other uncertainties in
the determination of extinction laws, this is probably a minor
contribution for the purposes of this work.

The opacity caused by a silicate resonance is defined as
(e.g. \citealt{DR03}):

\begin{equation}
\label{}
\Delta \tau_{\lambda} = \tau_{\lambda} - \tau_C,
\end{equation}

where $\tau_C$ is the optical depth of the ``continuum'' opacity
without the silicate resonance.  The relation between optical
extinction and $\Delta \tau$ is, using the fact that the optical depth
is proportional to the extinction coefficient $C^{\rm ext}$:

\begin{equation}
\label{X}
\frac{A_V}{\Delta \tau_{\lambda}} = \frac{\tau_{V}}{0.921 \Delta \tau_{\lambda}} = \frac{C^{\rm ext}_{V}}{0.921 \Delta C^{\rm ext}_{\lambda}}\equiv X_{\lambda} \,\rm mag.
\end{equation}

Using Equations \ref{es} and \ref{X} and by approximating $\delta
\rightarrow 0$, the silicate strength corrected for extinction is:

\begin{equation}
S_{\lambda}^{\rm corr} \sim \frac{S_{\lambda}}{\exp(-\Delta\tau_{\lambda})}
                        = \frac{S_{\lambda}}{\exp(-A_V/X_{\lambda})}
\end{equation}

\noindent
For the $R_V=5.5$ extinction law of \cite{WD01}, the relevant values
for $X_{\lambda}$ are 19.6, 20.9 and 41.6\,mag for 9.8, 10.0 and
11.3\,$\mu$m, respectively.

\begin{figure}[h!]
\begin{center}
\includegraphics[width=0.5\textwidth]{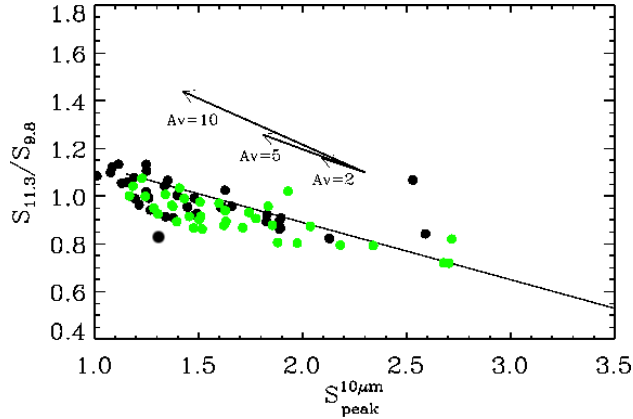}
\end{center}
\caption{\label{si_cor} Comparison between the observed values (in
black) and the extinction corrected values (in green) for the 10
$\mu$m silicate feature. The arrows give an indication of the effects
of extinction on the feature, however their length depends on the
starting point (see text for details). }
\end{figure}

Using this relationship, it is possible to correct the position of
sources with known values for $A_V$ in Figure \ref{si}. \citet{OL09}
measured $A_V$ values for 49 of the 89 disks studied here albeit with
some uncertainty. The resulting extinction-corrected strength-shape
distribution is shown in Figure \ref{si_cor} (in green) and compared
with the uncorrected distribution (in black). It is seen that
extinction moves points along curves that are almost parallel to the
relation between shape and strength. Most points, having $A_V<
5\,$mag, are not changed enough to alter the slope of the relation,
and the median strength is unaffected. However, a few highly extincted
disks are corrected by large amounts, as can be seen by the extinction
arrows in Figure \ref{si_cor}.

There are 40 YSOs in the sample for which \citet{OL09} do not estimate
$A_V$. Some of these may be highly extincted ($A_V \gtrsim 10$ mag)
sources (not being bright enough for optical spectroscopy using a 4-m
class telescope). The correction for extinction, as derived above,
would be greater for such objects, possibly introducing a displacement
between the distributions of extinction corrected and uncorrected
features. However, it is important to note that this assumes that the
$R_V=5.5$ extinction law is valid for the densest regions of molecular
clouds.  \citet{CH07} recently found that at $A_V>10\,$mag, the
relationship between optical depth and extinction derived for diffuse
ISM is no longer valid, possibly due to grain growth, and that $X_{\rm
9.7\,\mu m}$ is significantly larger in this regime. If true, then the
\cite{WD01} extinction law represents a ``worst case'' scenario. Since
most of the measured extinctions are small, and given the uncertainty
in extinction laws and the lack of $A_V$ data for a significant
fraction of the sample, the uncorrected silicate feature strengths are
used in the remainder of the paper.

\subsubsection{Statistical analysis of silicate features}
\label{ssi_stats}

\begin{figure}[h!]
\begin{center}
\includegraphics[width=0.5\textwidth]{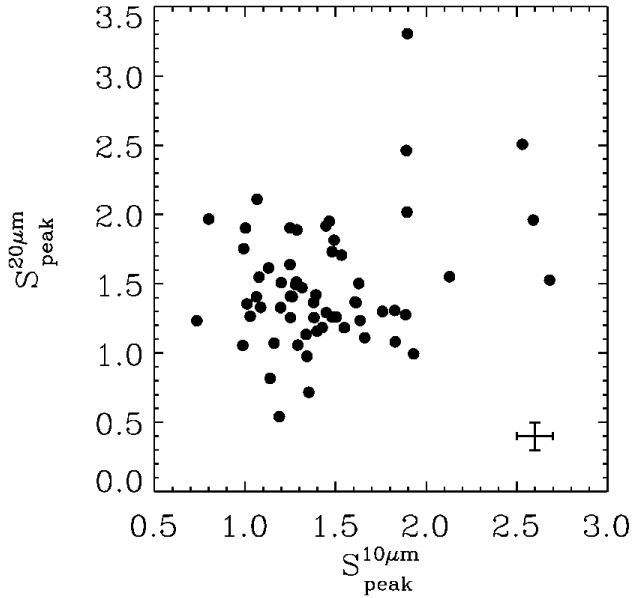}
\end{center}
\caption{\label{sipeak} Relative strength of the 10 vs. 20 $\mu$m
  silicate features. Typical error bars are shown in the bottom right
  corner. No obvious correlation is found. }
\end{figure}

Figure \ref{sipeak} shows the strengths of the observed 10 and 20
$\mu$m features with no apparent correlation. To quantify a
correlation, a Kendall $\tau$ rank correlation coefficient is used to
measure the degree of correspondence between two populations and to
assess the significance of said correspondence. In other words, if
there is a correlation (anti-correlation) between two datasets, the
Kendall $\tau$ rank coefficient is equal to 1 (-1). If the datasets
are completely independent, the coefficient has value 0. For the
strengths of the 10 and 20 $\mu$m silicate features, $\tau = 0.26$,
meaning at best a weak correlation.

The lack of apparent correlation between the 10 and the 20 $\mu$m
features, in both strength and appearance, suggests that they are
emitted by different regions in the disk. Indeed, several authors have
shown that the 10 $\mu$m feature is emitted by warm dust in the inner
region, while the 20 $\mu$m feature originates from a colder
component, further out and deeper into the disk (e.g.,
\citealt{KE07,ME07,BO08,OF09}). Therefore, the absence of the 10
$\mu$m feature but presence of the 20 $\mu$m band for a given source
implies that such disk lacks warmer small dust grains close to the
star. This reminisce the disk around the Herbig Ae star HD 100453
studied by \citet{ME02} and \citet{VA04}, who find that the absence of
the 10 $\mu$m silicate feature can be fitted with a model deprived in
warm small grains. 16\% of the disk sources shown in this sample fit
this scenario. A possible explanation for such a scenario is that
those disks have holes or gaps on such a small scale ($\lesssim$ 1 AU)
that the holes do not produce a strong signature in the spectra probed
by our data, as the cold disks do.

\section{Discussion}
\label{sdis}

In this section we discuss the properties of the YSO sample presented
in sections \ref{semb} and \ref{sdisks} with respect to environment,
and compare the results with those for another nearby star-forming
region, Taurus \citep{FU06}, as well as the full c2d IRS sample
(\citealt{KE06,OF09}).

\subsection{Cluster vs. Field Population}
\label{scf}

Comparison of the disk properties between the cluster and field
populations determines the importance of environment in the evolution
of these systems. If the evolution of the disks in clusters is found
to follow a different pace than that for more isolated stars, this can
set important constraints on disk evolution theories.

\begin{figure}[ht!]
\begin{center}
\includegraphics[width=0.5\textwidth]{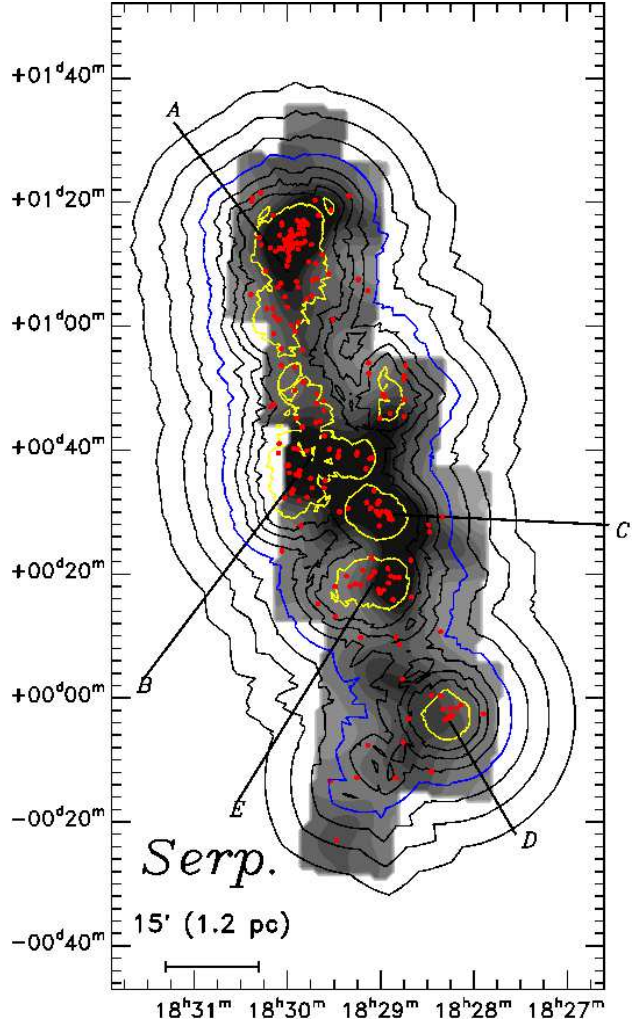}
\end{center}
\caption{\label{clusters} Clusters and groups in Serpens as defined
  according to the criteria in \citet{JG08}. The volume density
  contours are overlaid on the Serpens extinction map
  \citep{EV07}. The red dots are the YSOs in Serpens from
  \citet{HA07}. Black contours indicate volume densities of 0.125,
  0.25, 0.5, 2.0 and 4.0 M$_\odot$pc$^{-3}$, blue contour corresponds
  to volume density of 1 M$_\odot$pc$^{-3}$ and yellow contours to a
  volume density of 25 M$_\odot$pc$^{-3}$.}
\end{figure}

As discussed in \citet{HA06}, the brightest YSOs in Serpens appear to
be concentrated in clusters, but a more extended young stellar
population exists outside these clusters. For the determination of
clusters and their boundaries, we follow the method developed by
\citet{JG08} for Ophiuchus and Perseus. Using volume density and
number of YSOs as criteria, the method consists of a nearest neighbor
algorithm, and has as input the complete sample of YSOs in Serpens
\citep{HA07}. The associations are divided into loose (volume density
of 1 M$_\odot$pc$^{-3}$, blue contour in Figure \ref{clusters}) and
tight (volume density of 25 M$_\odot$pc$^{-3}$, yellow contours in
Figure \ref{clusters}). Furthermore, the associations are divided into
clusters (more than 35 members) and groups (less than 35 members).
The results of this method in Serpens yields 2 clusters (A and B) and
3 groups (C, D and E), that can be seen in Figure
\ref{clusters}. Individual memberships are marked in Table
\ref{tsi}. With the exception of one, virtually all YSOs in Serpens
are within the blue contour, i.e., all YSOs belong to, at least, a
loose association. This differs, for instance, from Ophiuchus and
Perseus \citep{JG08}, where some YSOs are completely isolated. For
number statistics of clustering in all c2d clouds, see \citet{EV09}.

Table \ref{tcluster} lists the number of objects in each tight
association. It appears that the ratio of disk to embedded sources in
a given cluster increases with distance from the densest part of the
cluster, where Group C is located (see Figure \ref{clusters}).

\begin{figure}[h!]
\begin{center}
\includegraphics[width=0.9\textwidth]{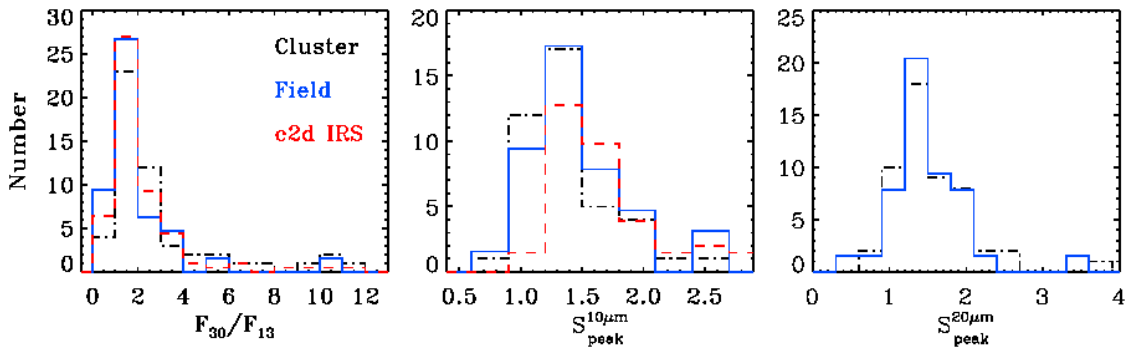}
\end{center}
\caption{\label{irscluster} Comparison between the clustered
  (dot-dashed black line) and field populations (solid blue line) of
  Serpens, with data from the c2d IRS program (dashed red line,
  \citealt{OF09}). In the left panel, the flux ratio between 30 and 13
  $\mu$m ($F_{30}/F_{13}$) is an indication of the disk geometry. The
  middle and right panels show the strength of the silicate features
  at 10 and 20 $\mu$m, respectively. }
\end{figure}

To compare the populations, all objects belonging to any of the tight
associations were grouped into the ``cluster population'', for better
number statistics.  Objects not belonging to any cluster were grouped
into the ``field population''.  Figure \ref{irscluster} shows the
comparison between these two populations for 3 quantities derived from
the IRS spectra.  In the left panel, the flux ratio between 30 and 13
$\mu$m ($F_{30}/F_{13}$) is an indication of disk geometry. The middle
and right panels show the peak intensity of the silicate features at
10 and 20 $\mu$m, respectively.  For all 3 quantities, the two
populations (cluster and field) are statistically indistinguishable. A
two sample Kolmogorov-Smirnov test (KS-test) was performed for each
quantity and the results show that the null hypothesis that the two
distributions come from the same parent population cannot be rejected
to any significance (18\%, 13\% and 79\% for $F_{30}/F_{13}$,
$S^{10\mu{\rm m}}_{{\rm peak}}$ and $S^{20\mu{\rm m}}_{{\rm peak}}$,
respectively).

Differences between the cluster vs. field populations will be further
investigated with ancillary data (e.g., relative stellar ages and
masses) and modeling (e.g., disk sizes). However, the IRS spectra
allow the conclusion that no significant differences are found for
disk geometry or the grain size distribution in the upper layers of
circumstellar disks in clustered compared with field stars. The latter
result only applies to the inner disk, as traced by silicate
features. The outer disk may still be different between cluster and
field populations.

\subsection{Comparison with Other Samples}
\label{sc2d}

The young stars with disks observed by the c2d team with IRS
spectroscopy (c2d IRS sample) are scattered across the sky in the 5
molecular clouds studied \citep{OF09}. The other 4 clouds studied by
the c2d are Chamaeleon II, Lupus, Perseus and Ophiuchus. All clouds
are nearby (within 300 pc), span a range of star-formation activity,
have typical median ages of a few Myr and have a spread between more
(Perseus and Ophiuchus) and less (Cha II and Lupus) clustered YSO
populations \citep{EV09}. The results from this sample are also
compared to the results in Serpens in Figures \ref{si} and
\ref{irscluster}. A conspicuous similarity is seen between the
samples. Note that \citet{OF09} do not analyze the 20 $\mu$m silicate
feature in this same manner and, therefore, $S^{20\mu{\rm m}}_{{\rm
    peak}}$ for the c2d IRS sample is missing from the right panel in
Figure \ref{irscluster}.

\begin{figure}[h!]
\begin{center}
\includegraphics[width=0.6\textwidth]{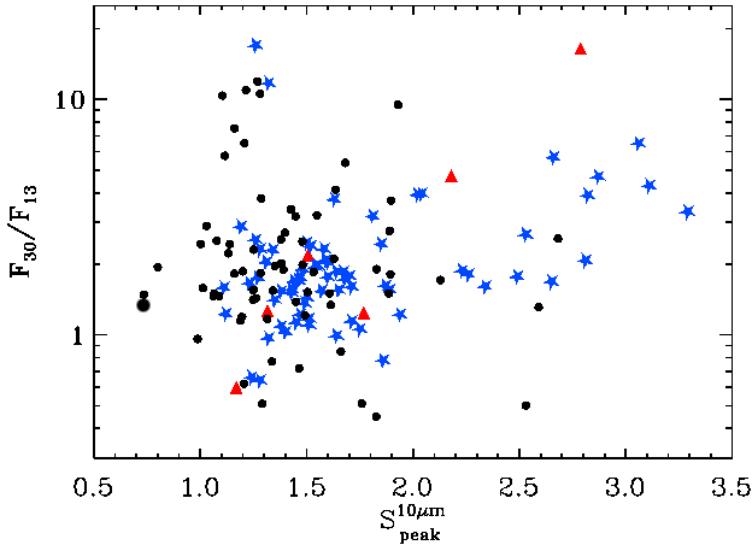}
\end{center}
\caption{\label{feps} The flux ratio between 30 and 13 $\mu$m
  ($F_{30}/F_{13}$) plotted against the peak at 10 $\mu$m
  ($S^{10\mu{\rm m}}_{{\rm peak}}$, black dots), compared against the
  T Tauri FEPS sample (red triangles, \citealt{BO08}) and the c2d IRS
  sample (blue stars, \citealt{OF09}). Typical uncertainties for the
  10 $\mu$m peak strength are $\sim$0.1, while typical uncertainties
  for the flux ratio are $\sim$0.12. }
\end{figure}

The models of \citet{DD08} conclude that, if sedimentation is the
unique reason for the variety of observed strength and shape of the 10
$\mu$m silicate feature, then this feature is strong for weak excess
in mid- to far-IR (as probed by $F_{30}/F_{13}$), and
vice-versa. However, when studying a small sample of T Tauri stars
from the Spitzer Legacy Program ``The Formation and Evolution of
Planetary Systems: Placing Our Solar System in Context'' (FEPS)
sample, \citet{BO08} found the opposite: a trend in which weak
$F_{30}/F_{13}$ correlates with a weak feature. A confirmation of this
trend for a larger sample implies that sedimentation alone cannot be
the sole cause for the diversity of observed silicate
features. \citet{DD08} argue that dust coagulation must then play a
vital role in producing different silicate profiles. In Figure
\ref{feps}, we populate this diagram with the Serpens sample (black
dots) and the c2d IRS sample (blue stars), as well as the Bouwman FEPS
sample (red triangles). This large combined sample shows no
correlation between the strength of the 10 $\mu$m silicate feature and
$F_{30}/F_{13}$ ($\tau = 0.07$) and therefore does not support either
the correlation (as seen by \citealt{BO08}) or the anti-correlation
(as modeled by \citet{DD08} for sedimentation alone) between the IR
flux excess and the strength of the 10 $\mu$m silicate feature.

\subsection{Comparison with Taurus}
\label{stau}

The Taurus Molecular Cloud is the best characterized star-forming
region to date, due to its proximity and relatively low
extinction. With young stars and their surrounding disks studied for
more than two decades (e.g., \citealt{KH87,KH95}), its members have
been well characterized at a wide range of wavelengths. Taurus has
thus become the reference for comparison of star-forming
regions. Compared to Serpens, Taurus seems to be a somewhat younger
(2.0 Myr median age, \citealt{LH01} vs. 4.7 Myr \citealt{OL09}) and has
lower star forming rate. Due to uncertainties in pre-main sequence age
determinations (e.g., \citealt{BA09,HI09,TN09}), this difference may
not be significant. However, cluster ages are statistically more
likely to be different than the same.

In a campaign similar to that presented here, IRS spectra were
obtained for a sample of 139 YSOs in Taurus, as part of a larger IRS
guaranteed-time observing program. These data were presented in two
separate papers: \citet{FU06} treated the disk sources, while
\citet{FU08} presented the embedded population in Taurus.

Because the young stellar population in Serpens has been discovered
and characterized using IR observations, this sample does not include
young stars without disks (and therefore no IR excess, also called
Class III sources, as defined in \citet{LA87}). For this reason, a
comparison between the Class III sources of the two clouds does not
make sense. 26 Class III objects were studied in Taurus. These objects
present featureless spectra with very little IR excess at longer
wavelengths. It is, however, possible to compare the IR excess
populations of both regions. Out of the entire YSO sample studied in
Taurus, embedded sources (or Class I) amount to 20\%{} while in
Serpens they amount to 18\%{}, a very comparable number. This also
matches the typical percentage of class I sources in the global YSO
statistics of the five clouds from the c2d photometric sample
\citep{EV09}. In Serpens only 5 objects have featureless spectra.

Excluding the embedded sources to analyze the disk population alone,
objects having both 10 and 20 $\mu$m silicate features amount to
72\%{} in Taurus, comparable to 73\%{} of the Serpens disk
population. Also comparable is the percentage of disks with PAH
emission: 3\% in Serpens and 4\% in Taurus.

In both regions, all disks with a 10 $\mu$m silicate feature also show
the 20 $\mu$m one. However, a difference is found for disks with only
the 20 $\mu$m feature in emission: they amount to 17\%{} of the disk
population in Serpens and only 4.5\%{} in Taurus. These statistics are
summarized in Table \ref{tstats}.

\begin{figure}[h!]
\begin{center}
\includegraphics[width=0.5\textwidth]{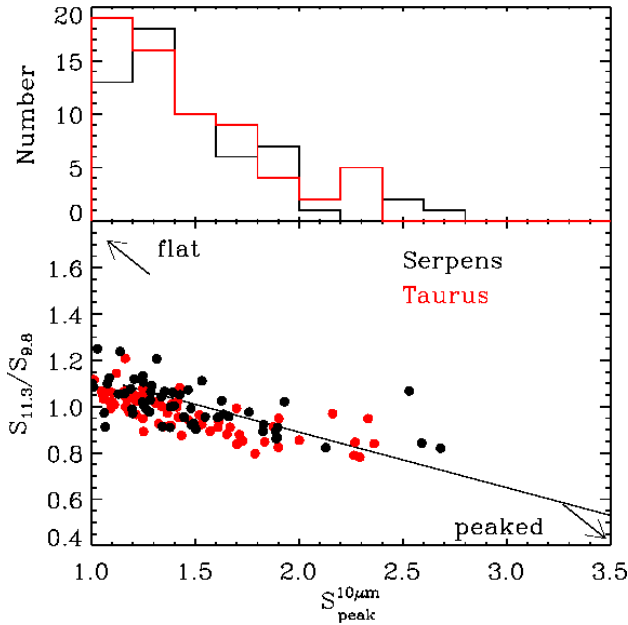}
\end{center}
\caption{\label{si_taurus} Top: Distribution of the 10 $\mu$m peak
  strength for Serpens (black) and Taurus (red). Bottom: Strength and
  shape of the 10 $\mu$m silicate feature for both Serpens (black) and
  Taurus (red).  }
\end{figure}

IRS data on the 85 disks sources presented in \citet{FU06} were
obtained from the Spitzer archive and reduced in the same manner as
our Serpens data. The same method described in Section \ref{ssi_ssr}
was applied to these data in order to compare the processes affecting
the dust in both regions. Figure \ref{si_taurus} (bottom) shows the
strength versus shape of the 10 $\mu$m silicate feature for both
Serpens (black dots) and Taurus (red dots), as well as the
distribution of peak strength for both regions (top). Serpens and
Taurus present remarkably similar distributions, with populations
clustered around flatter features and bigger grains, and almost no
strongly peaked silicate emission sources. A KS-test shows that the
hypothesis that the distributions are drawn from the same population
cannot be rejected (82\%).

\subsection{Implications}
\label{simp}

Comparisons between disks around stars that were formed in clusters or
in isolation in the Serpens Cloud, explored in \S~\ref{scf}, indicate
very similar populations. Thus, statistically no differences are found
in terms of both disk geometry (probed by the ratio $F_{30}/F_{13}$)
and processes affecting the dust (probed by the silicate
features). This suggests that local environment does not affect the
evolution path and timescale of disks.

Even more remarkable is the agreement between the YSO populations of
Serpens and Taurus, as well as the c2d IRS sample spread over 5
clouds, in terms of silicate features, as shown in \S~\ref{sc2d} and
\ref{stau}. Even though there are clear differences in silicate
features from source to source within a cloud, the overall {\it
  distribution} of feature shapes is statistically indistinguishable:
in all three samples, each containing of the order of 100 disks, the
bulk of the sources has a rather flat silicate profile, and a tail of
peaked shapes.

If the difference in median ages is significant, the similarity seen
in Figure \ref{si_taurus} indicates that a 2-3 Myr difference in age
is not reflected in a concurrent evolution of the average disk surface
dust properties. This indicates that the dust population in the disk
surface is an equilibrium between dust growth and destruction
processes, which is maintained at least as long as the disk is
optically thick. 

The process of grain growth through coagulation and settling to the
midplane has been shown to be much too short to be consistent with
disk observations \citep{WE80,DD05}. That means that the small grain
population must be replenished somehow, which could happen by
fragmentation of bigger aggregates and turbulent mixing. It is widely
accepted in the literature that the 10 $\mu$m silicate feature is
representative of the dust in the surface layers of the disk at a few
AU from the star \citep{KE07}. Significant evolution could still take
place in the disk mid-plane, which is not traced with these
observations. Indeed, millimeter observations have indicated that disk
midplanes are abundant in grains with mm/cm sizes
\citep{NA04b,RO06,LO07}. That means that, if this population of small
grains in the disk surface must be replenished by fragmentation of
bigger grains followed by turbulent mixing, the surface dust is an
indirect tracer of midplane grains.

If the age difference between Serpens and Taurus is significant, then
there exists a statistical equilibrium of processes of dust
coagulation and fragmentation in the disk lasting at least a few
million years that is independent of which YSO population is being
studied. The observable IR dust characteristics of a cluster
population of protoplanetary disks does not appear to depend on
cluster properties (age, density) within the first 5 Myr, for
relatively low-mass clusters. No specific property or event that may
influence the infrared dust signatures of a single disk (stellar
luminosity, spectral type, presence of a companion, disk-planet
interactions, disk instabilities etc.) produces detectable temporal or
spatial evolutionary effects visible in the distribution of disk
properties of an entire cluster within this time frame.

This is consistent with a picture in which a specific disk may change
its appearance on short time scales (much less than 1 Myr), but in a
{\it reversible} way. That is, the properties of disk surfaces may
oscillate between different states \citep{BA07,MU09}. If some
evolution of the disk surface is {\it irreversible}, and happens at a
given rate, a stable distribution of surface properties would not be
seen over time. This requires that the effect on the disk surface of
any reprocessing events has to be erased.

Another option is that the disk surface properties are determined by a
single parameter, such as the initial conditions of the formation of
such disks. In this scenario those properties should be kept
``frozen'' over the observed time scale of $\sim$few Myr. However,
this cannot be the case for objects like EX Lup, where real time
changes of the disk surface properties after the initial formation of
the disk have been observed on time scales of just a few years (namely
crystallization through thermal annealing of the dust in the disk
surface, \citealt{AB09}).

For this theory to work, the putative oscillation of states proposed
here must be stable over disk lifetimes of $\sim$5 Myr. As long as the
disks are gas-rich and optically thick, as in the three samples
studied here, we do not see a distinguishable evolution. It will be
important to search for surface evolution indicators in even older
clusters. Such comparisons will yield constraints on the global time
scale of disk evolution, which in turn will help constrain the
importance of the processes that play a vital role in the dissipation
of the disks.

The scenario in which a considerable population of small particles is
still present after bigger aggregates are formed and altered through
multiple events is consistent with evidence from primitive chondrites
in our own Solar System. Chondritic meteorites are observed to contain
fine-grained dust-like matrices formed after coarse-grained materials
such as chondrules and calcium-aluminum-rich inclusions (CAIs). This
indicates that the solar nebula underwent violent reprocessing events
in the feeding zones of parent bodies at 3-4 AU (see \citet{PB10} for
a recent review on Solar System dust in an astrophysical
context). These events included systemic melting of large dust
aggregates and possibly even evaporation and re-condensation of
silicate grains. In fact, presolar material is present in primitive
chondrites only at the trace level: $\sim$100 ppm \citep{LA05,ZI07},
testifying to the complete evaporative destruction of ISM dust at a
few AU in protoplanetary dust.

\section{Conclusions}
\label{scon}

We present Spitzer/IRS spectra from a complete and unbiased flux
limited sample of IR excess sources found in the Serpens Molecular
Cloud, following the c2d mapping of this region.

\begin{itemize}

\item Among our total of 147 IRS spectra, 22\% are found to be
  background contamination (including background stars, redshifted
  galaxies and a planetary nebula candidate). This high number is not
  surprising given the position of Serpens, close to the galactic
  plane.

\item Excluding the background objects from the sample, the bona fide
  set of YSOs amounts to 115 objects. The embedded to disk source
  ratio is 18\%, in agreement with the ratios derived from photometry
  \citep{EV09} for the five c2d clouds.

\item Disks with PAH in emission amount to 2\%\ of the YSO population,
  but 3\%\ of the disk population. Only G and A star show PAH
  emission. In the disk population, 73\%\ show both silicate emission
  features, at 10 and 20 $\mu$m, while 17\%\ show only the 20 $\mu$m
  feature. 4\%\ of the disks show featureless mid-IR spectra. Only one
  source, \#120, shows both PAH and silicate in emission.

\item Our YSO population in Serpens is very similar to that in Taurus,
  also studied with IRS spectra. In both regions about 70\%\ of the
  disk sources present both 10 and 20 $\mu$m silicate features in
  emission and the 10 $\mu$m feature is never seen without the 20
  $\mu$m feature. The only significant difference between the two
  populations is in the number of sources lacking the 10 $\mu$m
  feature but showing the 20 $\mu$m feature. This may be indicative of
  small holes ($\lesssim$ 1 AU) in these sources.

\item The silicate features in the IRS spectra measure the grain sizes
  that dominate the mid-IR opacity. The relationships between shape
  and strength of these features present distributions very similar to
  those obtained from other large samples of young stars.  Comparison
  with the models of \citet{OF09} yield grains consistent with sizes
  larger than a few $\mu$m.

\item No significant differences are found in the disk geometry or the
  grain size distribution in the upper layers of circumstellar disks
  (probed by the silicate features) in clustered or field stars in the
  cloud, indicating that the local environment where a star is born
  does not influence the evolution of its harboring disk.

\item Quantitatively, the shape and strength of the 10 $\mu$m silicate
  feature were used to compare both Serpens and Taurus as well as a
  large sample of disks across five clouds, indicating remarkably
  similar populations. This implies that the dust population in the
  disk surface results from an equilibrium between dust growth and
  destruction processes that are maintained over a few million years.

\end{itemize}

\acknowledgements Support for this work, part of the Spitzer Space
Telescope Legacy Science Program, was provided by NASA through
Contract Numbers 1256316, 1224608 and 1230780 issued by the Jet
Propulsion Laboratory, California Institute of Technology under NASA
contract 1407 and by the Spanish Grant AYA 2005-0954. Astrochemistry
at Leiden is supported by a Spinoza grant from the Netherlands
Organization for Scientific Research (NWO) and by the Netherlands
Research School for Astronomy (NOVA) grants. Support to KMP was
provided by NASA through Hubble Fellowship grant \#01201.01, awarded
by the Space Telescope Science Institute, which is operated by the
Association of Universities for Research in Astronomy, Inc., for NASA,
under contract NAS 5-26555.  The authors are very grateful to
L. Allamandola, R. Overzier, P. Beir\~ao, R. Demarco, J. Green, and
A. K\'ospal for fruitful discussions, and would like to acknowledge
the anonymous referee for suggestions that have improved the
manuscript considerably.

\vspace{5cm}
{\it Facilities:} \facility{Spitzer/IRS}

\clearpage

\begin{deluxetable}{r r c c c r l c c }
\tabletypesize{\scriptsize}
\tablecolumns{9}
\tablewidth{0pt}
\tablecaption{Observed objects in Serpens \label{tab1}} 
\tablehead{\colhead{\#\tablenotemark{a}}  & 
           \colhead{ID\tablenotemark{b}}  &  
           \colhead{c2d ID (SSTc2dJ)}     &  
           \colhead{$\alpha_{2\mu{\rm m}-24\mu{\rm m}}$\tablenotemark{c}}  &
           \colhead{Program\tablenotemark{d}}                          &
           \colhead{AOR}                  &
           \colhead{Obs. Date}            &
           \colhead{$F_{30}/F_{13}$\tablenotemark{e}}                   &
           \colhead{Classification}
}
\startdata
  1 &   1 &  18275383-0002335  & $-$1.42  & GO3  & 17891072  & 2007-04-19 &  1.31  & Disk          \\
  2 &   2 &  18280503+0006591  & $-$2.29  & GO3  & 17883648  & 2007-04-24 &  0.35  & BG star       \\
  3 &   3 &  18280845-0001064  & $-$0.70  & GO3  & 17883136  & 2007-04-24 &  1.52  & Disk          \\
  4 &   4 &  18281100-0001395  & $-$2.13  & GO3  & 17888256  & 2007-04-29 &  0.27  & BG star       \\
  5 &     &  18281315+0003128  & $-$1.84  & c2d  & 13210368  & 2005-05-20 &  0.27  & BG star       \\
  6 &   5 &  18281350-0002491  & $-$1.08  & GO3  & 17883136  & 2007-04-24 &  1.71  & Disk          \\
  7 &   6 &  18281501-0002588  & $-$0.05  & GO3  & 17883136  & 2007-04-24 &  2.54  & Disk          \\
  8 &   7 &  18281519-0001407  & $-$1.25  & GO3  & 17889536  & 2007-04-30 &  2.43  & Disk          \\
  9 &   8 &  18281525-0002434  & $-$0.85  & GO3  & 17882368  & 2007-04-25 &  5.37  & Cold disk     \\
 10 &   9 &  18281629-0003164  & $-$1.13  & GO3  & 17884160  & 2007-04-19 &  1.83  & Disk          \\
 11 &     &  18281757-0006474  &    0.65  & GO3  & 17890560  & 2007-04-28 &  3.36  & Galaxy        \\
 12 &     &  18281757+0016065  &    0.53  & GO3  & 17888768  & 2007-05-05 &  4.72  & Galaxy        \\
 13 &  11 &  18281981-0001475  & $-$0.89  & GO3  & 17889792  & 2007-04-30 &  4.09  & Disk          \\
 14 &  13 &  18282143+0010411  & $-$1.44  & GO3  & 17888000  & 2007-04-30 &  1.38  & Disk          \\
 15 &  14 &  18282159+0000162  & $-$0.99  & GO3  & 17883136  & 2007-04-24 &  2.56  & Disk          \\
 16 &  15 &  18282432+0034545  & $-$2.01  & GO3  & 17887744  & 2007-04-24 &  0.32  & BG star       \\
 17 &     &  18282720+0044450  &    1.82  & GO3  & 17884928  & 2007-04-29 & 11.04  & PN candidate \\
 18 &  16 &  18282738-0011499  & $-$1.98  & GO3  & 17883136  & 2007-04-24 &  0.32  & BG star       \\
 19 &  17 &  18282741+0000239  & $-$1.97  & GO3  & 17887744  & 2007-04-24 &  0.43  & BG star       \\
 20 &  18 &  18282849+0026500  & $-$1.12  & GO3  & 17890048  & 2007-04-29 &  1.13  & Disk          \\
 21 &  19 &  18282905+0027560  & $-$1.06  & GO3  & 17884928  & 2007-04-29 & 10.38  & Cold disk     \\
 22 &     &  18283000+0020147  & $-$1.60  & c2d  & 13210624  & 2005-05-20 &  0.29  & BG star       \\
 23 &     &  18283736+0019276  & $-$2.19  & GO3  & 17887744  & 2007-04-24 &  0.27  & BG star       \\
 24 &  20 &  18284025+0016173  & $-$0.96  & GO3  & 17890048  & 2007-04-29 &  1.20  & Disk          \\
 25 &  21 &  18284053+0022144  & $-$1.91  & GO3  & 17887744  & 2007-04-24 &  0.42  & BG star       \\
 26 &  22 &  18284187-0003215  &    0.53  & GO3  & 17883136  & 2007-04-24 &  3.39  & Emb           \\
 27 &  23 &  18284403+0053379  &    0.70  & GO3  & 17882880  & 2007-04-28 &  3.37  & Emb           \\
 28 &  24 &  18284479+0051257  &    1.04  & GO3  & 17885184  & 2007-04-24 & 10.18  & Emb           \\
 29 &  25 &  18284481+0048085  & $-$1.01  & GO3  & 17888512  & 2007-05-05 &  2.76  & Disk          \\
 30 &  27 &  18284497+0045239  & $-$1.35  & GO3  & 17886720  & 2007-04-24 &  0.77  & Disk          \\
 31 &  38 &  18284559-0007132  & $-$0.77  & GO3  & 17882368  & 2007-04-25 &  0.51  & Disk          \\
 32 &  29 &  18284614+0003016  & $-$0.64  & GO3  & 17889024  & 2007-04-25 &  0.66  & Disk          \\
 33 &     &  18284632-0011103  &    0.53  & GO3  & 17890560  & 2007-04-28 & 11.17  & Galaxy        \\
 34 &     &  18284828-0005300  & $-$2.43  & GO3  & 17888256  & 2007-04-29 &  0.47  & BG star       \\
 35 &     &  18284938-0006046  & $-$2.01  & c2d  & 13210624  & 2005-05-20 &  0.28  & BG star       \\
 36 &  32 &  18285020+0009497  & $-$0.24  & c2d  & 13461505  & 2006-04-31 &  1.98  & Disk          \\
 37 &  33 &  18285039-0012552  & $-$1.40  & GO3  & 17884160  & 2007-04-19 &  0.32  & BG star       \\
 38 &     &  18285060+0007540  & $-$2.31  & c2d  & 13460736  & 2005-09-09 &   --   & Disk          \\
 39 &  34 &  18285122+0019271  &    0.45  & GO3  & 17883904  & 2007-04-29 &  8.98  & Emb           \\
 40 &  36 &  18285249+0020260  & $-$0.15  & GO3  & 17883136  & 2007-04-24 &  1.46  & Disk          \\
 41 &  37 &  18285276+0028466  &    0.07  & GO3  & 17887232  & 2007-04-27 &  9.50  & Disk          \\
 42 &  38 &  18285362+0019302  & $-$0.64  & GO3  & 17887488  & 2007-04-25 &  2.09  & Disk          \\
 43 &  39 &  18285395+0045530  & $-$1.16  & GO3  & 17886208  & 2007-09-30 &  1.96  & Disk          \\
 44 &  40 &  18285404+0029299  &    1.35  & c2d  & 13461249  & 2006-04-20 & 18.21  & Emb           \\
 45 &  42 &  18285450+0029520  &    0.15  & c2d  & 13460993  & 2006-04-19 & 35.49  & Emb           \\
 47 &  43 &  18285489+0018326  &    0.88  & GO3  & 17883392  & 2007-04-29 &  7.58  & Emb           \\
 48 &  44 &  18285529+0020522  & $-$0.29  & GO3  & 17884160  & 2007-04-19 &  1.43  & Disk          \\
 49 &  45 &  18285580+0029444  &    1.81  & GO3  & 17883392  & 2007-04-29 & 99.75  & Emb           \\
 50 &  46 &  18285660+0030080  &    1.84  & GO3  & 17889792  & 2007-04-30 & 21.37  & Emb           \\
 51 &  47 &  18285719+0048359  & $-$0.32  & GO3  & 17888768  & 2007-05-05 & 11.27  & Emb           \\
 52 &  48 &  18285808+0017244  & $-$2.11  & GO3  & 17888000  & 2007-04-30 &  1.23  & Disk          \\
 53 &  49 &  18285860+0048594  & $-$1.06  & GO3  & 17887488  & 2007-04-25 &  1.55  & Disk          \\
 54 &  50 &  18285946+0030029  & $-$0.59  & c2d  & 13461249  & 2006-04-20 &  0.62  & Disk          \\
 55 &  51 &  18290025+0016580  & $-$1.01  & GO3  & 17884416  & 2007-04-30 &  1.50  & Disk          \\
 56 &  52 &  18290057+0045079  & $-$0.94  & GO3  & 17886208  & 2007-09-30 &  3.22  & Disk          \\
 57 &  53 &  18290082+0027467  & $-$0.36  & GO3  & 17885440  & 2007-04-24 &  3.41  & Disk          \\
 58 &  54 &  18290088+0029315  & $-$0.37  & c2d  & 13210112  & 2005-04-17 &  1.82  & Disk          \\
 59 &  55 &  18290107+0031451  & $-$0.49  & c2d  & 13461249  & 2006-04-20 &  1.14  & Disk          \\
 60 &  56 &  18290122+0029330  & $-$0.39  & c2d  & 13461505  & 2006-04-21 &  1.58  & Disk          \\
 61 &  58 &  18290175+0029465  & $-$0.84  & c2d  & 13461249  & 2006-04-20 &  2.30  & Disk          \\
 62 &  59 &  18290184+0029546  & $-$0.77  & c2d  & 13210112  & 2005-04-17 &  1.32  & Disk          \\
 63 &  60 &  18290210+0031210  &    0.29  & GO3  & 17887232  & 2007-04-27 & 32.23  & Emb           \\
 64 &     &  18290215+0029400  & $-$0.86  & GO3  & 17885184  & 2007-04-24 &  2.89  & Disk          \\
 65 &  61 &  18290286+0030089  & $-$0.23  & c2d  & 13460993  & 2006-04-21 &  2.43  & Disk          \\
 66 &  62 &  18290393+0020217  & $-$0.82  & GO3  & 17883136  & 2007-04-24 &  1.19  & Disk          \\
 67 &  63 &  18290436+0033237  &    0.95  & c2d  & 13461505  & 2006-06-21 &  2.54  & Emb           \\
 68 &     &  18290442-0020018  &    0.52  & GO3  & 17890560  & 2007-04-28 &  5.61  & Galaxy        \\
 69 &  64 &  18290518+0038438  & $-$1.24  & GO3  & 17889792  & 2007-04-30 &  5.64  & Cold disk     \\
 70 &  65 &  18290575+0022325  & $-$2.12  & GO3  & 17886976  & 2007-04-29 &  1.76  & Disk          \\
 71 &  66 &  18290615+0019444  & $-$1.52  & GO3  & 17888512  & 2007-05-05 &  2.49  & Disk          \\
 72 &  67 &  18290620+0030430  &    1.69  & c2d  & 13461249  & 2006-04-21 & 17.07  & Emb           \\
 73 &  68 &  18290680+0030340  &    1.67  & c2d  & 13460993  & 2006-04-21 & 75.97  & Emb           \\
 74 &  69 &  18290699+0038377  & $-$0.63  & GO3  & 17884416  & 2007-04-30 &  2.90  & Disk          \\
 75 &  71 &  18290765+0052223  & $-$1.01  & GO3  & 17888512  & 2007-05-05 &  2.01  & Disk          \\
 76 &  72 &  18290775+0054037  & $-$1.06  & GO3  & 17888000  & 2007-04-30 &  1.34  & Disk          \\
 77 &  73 &  18290808-0007371  & $-$2.22  & GO3  & 17888256  & 2007-04-29 &  0.27  & BG star       \\
 78 &     &  18290843-0026187  & $-$2.28  & GO3  & 17885696  & 2007-04-28 &  0.31  & BG star       \\
 79 &  75 &  18290910+0031300  &    0.24  & GO3  & 17884928  & 2007-04-29 & 51.02  & Emb           \\
 80 &  76 &  18290956+0037016  & $-$1.13  & GO3  & 17882368  & 2007-04-25 &  0.51  & Disk          \\
 81 &  77 &  18290980+0034459  & $-$0.97  & c2d  & 13210624  & 2005-05-20 &  0.85  & Disk          \\
 82 &  78 &  18291148+0020387  & $-$1.34  & GO3  & 17887232  & 2007-04-27 & 11.90  & Cold disk     \\
 83 &  79 &  18291249+0018152  & $-$1.06  & GO3  & 17889536  & 2007-04-30 &  2.06  & Disk          \\
 84 &  81 &  18291407+0002589  & $-$2.19  & GO3  & 17885696  & 2007-04-28 &  0.30  & BG star       \\
 85 &     &  18291477-0004237  & $-$2.29  & c2d  & 13210112  & 2005-04-17 &  0.24  & BG star       \\
 86 &  83 &  18291508+0052124  & $-$1.89  & GO3  & 17882880  & 2007-04-28 &  0.45  & Disk          \\
 87 &  84 &  18291513+0039378  & $-$1.50  & GO3  & 17889024  & 2007-04-25 &  1.01  & Disk          \\
 88 &  85 &  18291539-0012519  & $-$1.72  & GO3  & 17885952  & 2006-10-23 &  0.50  & Disk          \\
 89 &  86 &  18291557+0039119  & $-$0.56  & GO3  & 17884416  & 2007-04-30 &  2.51  & Disk          \\
 90 &  87 &  18291563+0039228  & $-$0.89  & GO3  & 17885184  & 2007-04-24 &  6.52  & Cold disk     \\
 91 &  88 &  18291617+0018227  &    0.45  & c2d  & 13210112  & 2005-04-17 &  2.11  & Emb           \\
 92 &  89 &  18291969+0018031  & $-$1.23  & GO3  & 17885440  & 2007-04-24 &  2.10  & Disk          \\
 93 &  90 &  18292001+0024497  & $-$2.04  & GO3  & 17886720  & 2007-04-24 &  0.29  & BG star       \\
 94 &     &  18292050+0047080  &    1.29  & GO3  & 17888768  & 2007-05-05 & 25.76  & Disk          \\
 95 &  93 &  18292094+0030345  & $-$1.73  & GO3  & 17882880  & 2007-04-28 &  0.40  & BG star       \\
 96 &  94 &  18292184+0019386  & $-$0.99  & GO3  & 17886976  & 2007-04-29 &  1.54  & Disk          \\
 97 &     &  18292250+0034118  & $-$1.65  & GO3  & 17889280  & 2007-04-24 &  0.29  & Disk          \\
 98 &     &  18292253+0034176  & $-$2.48  & GO3  & 17889024  & 2007-04-25 &  1.34  & Disk          \\
 99 &  95 &  18292616+0020518  & $-$2.26  & GO3  & 17883648  & 2007-04-24 &  0.37  & BG star       \\
100 &  96 &  18292640+0030042  & $-$0.98  & GO3  & 17888768  & 2007-05-05 &  1.89  & Disk          \\
101 &  97 &  18292679+0039497  & $-$0.60  & GO3  & 17885440  & 2007-04-24 &  4.13  & Disk          \\
102 &  99 &  18292736+0038496  &    0.29  & GO3  & 17886464  & 2007-04-29 &  3.95  & Emb           \\
103 & 100 &  18292824-0022574  & $-$0.41  & c2d  & 13210368  & 2005-05-20 &  1.17  & Disk          \\
104 &     &  18292833+0049569  & $-$0.48  & GO3  & 17887488  & 2007-04-25 &  1.85  & Disk          \\
105 &     &  18292864+0042369  & $-$0.97  & GO3  & 17890048  & 2007-04-29 &  1.94  & Disk          \\
106 & 102 &  18292927+0018000  & $-$1.24  & GO3  & 17889792  & 2007-04-30 &  1.15  & Disk          \\
107 &     &  18293056+0033377  & $-$0.60  & GO3  & 17889536  & 2007-04-30 &  1.46  & Disk          \\
108 & 105 &  18293254-0013233  & $-$2.58  & GO3  & 17884672  & 2007-04-28 &  1.36  & BG star       \\
109 & 106 &  18293300+0040087  & $-$1.35  & GO3  & 17888512  & 2007-05-05 &  2.16  & Disk          \\
110 & 107 &  18293319+0012122  & $-$1.81  & GO3  & 17883648  & 2007-04-24 &  0.34  & BG star       \\
111 &     &  18293337+0050136  & $-$0.26  & GO3  & 17888768  & 2007-05-05 &  3.80  & Disk          \\
112 & 109 &  18293381+0053118  & $-$2.06  & GO3  & 17886720  & 2007-04-24 &  0.29  & BG star       \\
113 & 110 &  18293561+0035038  & $-$1.53  & GO3  & 17883904  & 2007-04-29 &  5.76  & Cold disk     \\
114 & 111 &  18293619+0042167  & $-$0.92  & GO3  & 17883904  & 2007-04-29 &  7.52  & Cold disk     \\
115 & 114 &  18293672+0047579  & $-$1.00  & GO3  & 17886976  & 2007-04-29 &  3.19  & Disk          \\
116 & 116 &  18293882+0044380  & $-$1.23  & GO3  & 17890048  & 2007-04-29 &  1.55  & Disk          \\
117 & 118 &  18294020+0015131  &    0.87  & GO3  & 17883392  & 2007-04-29 &  2.72  & Disk          \\
118 &     &  18294067-0007033  & $-$2.53  & GO3  & 17884672  & 2007-04-28 &  0.38  & BG star       \\
119 & 119 &  18294121+0049020  & $-$1.39  & GO3  & 17889280  & 2007-04-24 &  1.55  & Disk          \\
120 & 123 &  18294168+0044270  & $-$1.42  & GO3  & 17884416  & 2007-04-30 &  3.13  & Disk          \\
121 &     &  18294301-0016083  & $-$2.18  & GO3  & 17890304  & 2006-10-25 &  0.33  & BG star       \\
122 & 126 &  18294410+0033561  & $-$1.68  & GO3  & 17886464  & 2007-04-29 & 10.92  & Cold disk     \\
123 & 129 &  18294503+0035266  & $-$1.26  & GO3  & 17886976  & 2007-04-29 &  3.71  & Disk          \\
124 & 132 &  18294725+0039556  & $-$1.32  & GO3  & 17887232  & 2007-04-27 &  2.22  & Disk          \\
125 & 133 &  18294726+0032230  & $-$1.31  & GO3  & 17886976  & 2007-04-29 &  1.50  & Disk          \\
126 & 140 &  18294962+0050528  & $-$2.00  & GO3  & 17890816  & 2007-04-24 &  0.35  & BG star       \\
127 & 143 &  18295001+0051015  & $-$1.68  & GO3  & 17889280  & 2007-04-24 &  1.23  & Disk          \\
128 &     &  18295012+0027229  &    0.14  & GO3  & 17888768  & 2007-05-05 &  2.60  & Galaxy        \\
129 & 144 &  18295016+0056081  & $-$0.76  & GO3  & 17886208  & 2007-09-30 &  1.86  & Disk          \\
130 & 145 &  18295041+0043437  & $-$1.32  & GO3  & 17886720  & 2007-04-24 &  2.14  & Disk          \\
131 & 148 &  18295130+0027479  & $-$2.42  & GO3  & 17889024  & 2007-04-25 &  2.09  & Disk          \\
132 & 149 &  18295206+0036436  & $-$0.06  & GO3  & 17886464  & 2007-04-29 &  4.04  & Emb           \\
133 &     &  18295240+0035527  & $-$0.33  & GO3  & 17884416  & 2007-04-30 &  3.07  & Emb           \\
134 & 153 &  18295244+0031496  & $-$0.59  & GO3  & 17889792  & 2007-04-30 &  1.48  & Disk          \\
135 & 154 &  18295252+0036116  &    0.78  & GO3  & 17883392  & 2007-04-29 & 24.39  & Emb           \\
136 & 156 &  18295304+0040105  & $-$1.16  & GO3  & 17889536  & 2007-04-30 &  1.50  & Disk          \\
137 & 157 &  18295305+0036065  &    0.07  & GO3  & 17882624  & 2006-10-23 & 10.57  & Disk          \\
138 & 161 &  18295322+0033129  & $-$1.99  & GO3  & 17886720  & 2007-04-24 &  2.35  & BG star       \\
139 & 165 &  18295422+0045076  & $-$2.35  & GO3  & 17889024  & 2007-04-25 &  1.03  & Disk          \\
140 & 166 &  18295435+0036015  & $-$0.19  & GO3  & 17887232  & 2007-04-27 &  9.65  & Emb           \\
142 & 172 &  18295592+0040150  & $-$0.85  & GO3  & 17888000  & 2007-04-30 &  1.21  & Disk          \\
143 & 173 &  18295620+0033391  & $-$0.54  & GO3  & 17889024  & 2007-04-25 &  1.21  & Disk          \\
144 & 177 &  18295701+0033001  & $-$1.15  & GO3  & 17882880  & 2007-04-28 &  0.96  & Disk          \\
145 & 178 &  18295714+0033185  & $-$0.82  & GO3  & 17883648  & 2007-04-24 &  1.26  & Disk          \\
146 & 182 &  18295772+0114057  & $-$0.44  & c2d  & 9828352   & 2004-09-02 &  1.41  & Disk          \\
147 & 189 &  18295872+0036205  & $-$1.11  & GO3  & 17888000  & 2007-04-30 &  0.72  & Disk          \\
148 & 206 &  18300178+0032162  & $-$1.25  & GO3  & 17886976  & 2007-04-29 &  1.90  & Disk          \\
149 & 210 &  18300350+0023450  & $-$1.22  & GO3  & 17888000  & 2007-04-30 &  1.81  & Disk          \\
\enddata
\tablenotetext{a}{As in \citet{OL09}}
\tablenotetext{b}{From \citet{HA07}}
\tablenotetext{c}{$\frac{d \log(\lambda F_\lambda)}{d \log(\lambda)}$ between the 2MASS $K$ (2$\mu$m) and the MIPS1 (24$\mu$m) bands \citep{HA07}}
\tablenotetext{d}{`GO3' means the object was observed as part of the Spitzer Space Telescope's cycle 3 program \#30223, PI: K. Pontoppidan; `c2d' means the object was observed as part of the c2d 2nd Look program, PI: N. Evans}
\tablenotetext{e}{Flux ratio between 30 and 13 $\mu$m}
\end{deluxetable}

\setlength{\voffset}{0mm}

\begin{deluxetable}{r c c c c c c c c c c }
\tabletypesize{\footnotesize}
\tablecolumns{11}
\tablewidth{0pt}
\tablecaption{Characteristics of YSOs in Serpens \label{tsi}} 
\tablehead{\colhead{\#}                      & 
           \colhead{Emb}                     &
           \colhead{PAH\tablenotemark{a}}    &
           \colhead{S$_{9.8}$}                &
           \colhead{S$_{11.3}$}               &
           \colhead{S$^{10\mu{}m}_{{\rm peak}}$} &
           \colhead{S$_{19.0}$}               &
           \colhead{S$_{23.8}$}               &
           \colhead{S$^{20\mu{}m}_{{\rm peak}}$} &
           \colhead{Cluster}                 & 
           \colhead{Spectral Type}      
}
\startdata
  1   &  -- & -- &  2.53  &  2.13  &  2.59  &  2.00  &  1.52  &  1.96  &     & K2  \\
  3   &  -- & -- &  1.49  &  1.35  &  1.50  &  1.33  &  1.23  &  1.26  &  D  & M0  \\
  6   &  -- & -- &  2.12  &  1.73  &  2.13  &  1.61  &  1.33  &  1.55  &  D  & K5  \\
  7   &  -- & -- &  1.38  &  1.25  &  1.38  &  1.39  &  1.34  &  1.36  &  D  & M0  \\
  8   &  -- & -- &  1.15  &  1.43  &  1.14  &  1.21  &  1.45  &  0.82  &  D  &   \\
  9   &  -- & -- &  1.64  &  1.48  &  1.68  &  2.24  &  1.97  &  2.27  &  D  &   \\
 10   &  -- & -- &  1.25  &  1.29  &  1.28  &  1.47  &  1.29  &  1.51  &  D  &   \\
 13   &  -- & -- &   --   &   --   &   --   &  1.33  &  1.87  &  1.29  &  D  &   \\
 14   &  -- & -- &  1.45  &  1.38  &  1.45  &  1.65  &  1.43  &  1.53  &     & M2  \\
 15   &  -- & -- &  2.67  &  2.19  &  2.68  &  2.59  &  2.07  &  2.46  &  D  &   \\
 20   &  -- & -- &   --   &   --   &   --   &  0.90  &  0.85  &  1.13  &     &   \\
 21   &  -- & -- &  1.09  &  1.17  &  1.10  &  1.45  &  1.57  &  1.34  &     &   \\
 24   &  -- & -- &   --   &   --   &   --   &  1.69  &  1.29  &  1.30  &     &   \\
 26   &  Y  & -- &   --   &   --   &   --   &   --   &   --   &   --   &     &   \\
 27   &  Y  & -- &   --   &   --   &   --   &   --   &   --   &   --   &     &   \\
 28   &  Y  & -- &   --   &   --   &   --   &   --   &   --   &   --   &     &   \\
 29   &  -- & -- &  1.84  &  1.59  &  1.89  &  1.56  &  2.03  &  1.73  &     & M2  \\
 30   &  -- & -- &  1.32  &  1.38  &  1.34  &  1.22  &  1.18  &  1.33  &     & M1  \\
 31   &  -- & -- &  1.76  &  1.72  &  1.76  &  0.99  &  0.92  &  0.99  &     &   \\
 32   &  -- & -- &   --   &   --   &   --   &  0.89  &  1.13  &  0.72  &     &   \\
 36   &  -- & -- &  1.47  &  1.35  &  1.48  &  1.41  &  1.31  &  1.41  &     & K5  \\
 38   &  -- & -- &   --   &   --   &   --   &   --   &   --   &   --   &     &   \\
 39   &  Y  & -- &   --   &   --   &   --   &   --   &   --   &   --   &  E  &   \\
 40   &  -- & -- &  1.06  &  1.20  &  1.09  &  1.29  &  1.26  &  1.25  &  E  & M7  \\
 41   &  -- & -- &  1.73  &  1.77  &  1.93  &  3.45  &  4.18  &  3.74  &  C  & K2  \\
 42   &  -- & -- &   --   &   --   &   --   &   --   &   --   &   --   &  E  &   \\
 43   &  -- & -- &  1.32  &  1.40  &  1.35  &  1.18  &  1.18  &  1.28  &     & M0.5  \\
 44   &  Y  & -- &   --   &   --   &   --   &   --   &   --   &   --   &  C  & M6  \\
 45   &  Y  & -- &   --   &   --   &   --   &   --   &   --   &   --   &  C  &   \\
 47   &  Y  & -- &   --   &   --   &   --   &   --   &   --   &   --   &  E  & M5  \\
 48   &  -- & -- &  1.22  &  1.21  &  1.26  &  1.12  &  1.05  &  1.18  &  E  & M5.5  \\
 49   &  Y  & -- &   --   &   --   &   --   &   --   &   --   &   --   &  C  &   \\
 50   &  Y  & -- &   --   &   --   &   --   &   --   &   --   &   --   &  C  &   \\
 51   &  Y  & -- &   --   &   --   &   --   &   --   &   --   &   --   &     &   \\
 52   &  -- & Y  &   --   &   --   &   --   &   --   &   --   &   --   &  E  & G3  \\
 53   &  -- & -- &  1.19  &  1.32  &  1.25  &  1.26  &  1.04  &  1.18  &     & M2.5  \\
 54   &  -- & -- &  1.13  &  1.26  &  1.21  &  1.19  &  1.47  &  1.07  &  C  &   \\
 55   &  -- & -- &  1.88  &  1.68  &  1.89  &  1.53  &  1.18  &  1.35  &  E  & K2  \\
 56   &  -- & -- &  1.56  &  1.49  &  1.55  &  1.32  &  1.23  &  1.41  &     &   \\
 57   &  -- & -- &  1.39  &  1.47  &  1.43  &  2.01  &  1.81  &  1.90  &  C  &   \\
 58   &  -- & -- &  1.09  &  1.15  &  1.16  &  1.13  &  1.20  &  1.33  &  C  & K7  \\
 59   &  -- & -- &   --   &   --   &   --   &   --   &   --   &   --   &  C  &   \\
 60   &  -- & -- &  1.00  &  1.08  &  1.01  &  1.27  &  1.25  &  1.26  &  C  & M0.5  \\
 61   &  -- & -- &  1.24  &  1.25  &  1.25  &  1.28  &  1.32  &  1.26  &  C  & M0  \\
 62   &  -- & -- &   --   &   --   &   --   &  1.21  &  1.31  &  1.26  &  C  & K0  \\
 63   &  Y  & -- &   --   &   --   &   --   &   --   &   --   &   --   &  C  &   \\
 64   &  -- & -- &   --   &   --   &   --   &  1.41  &  1.27  &  1.36  &  C  &   \\
 65   &  -- & -- &  1.00  &  1.10  &  1.00  &  1.21  &  1.12  &  1.06  &  C  &   \\
 66   &  -- & -- &  1.19  &  1.18  &  1.20  &  1.15  &  1.07  &  1.11  &  E  & K5  \\
 67   &  Y  & -- &   --   &   --   &   --   &   --   &   --   &   --   &  C  &   \\
 69   &  -- & -- &   --   &   --   &   --   &  2.14  &  2.17  &  2.27  &  B  &   \\
 70   &  -- & -- &   --   &   --   &   --   &  1.29  &  1.36  &  1.31  &  E  & A3  \\
 71   &  -- & -- &  1.50  &  1.49  &  1.48  &  2.48  &  2.86  &  2.51  &  E  & M3  \\
 72   &  Y  & -- &   --   &   --   &   --   &   --   &   --   &   --   &  C  &   \\
 73   &  Y  & -- &   --   &   --   &   --   &   --   &   --   &   --   &  C  &   \\
 74   &  -- & -- &  0.99  &  1.24  &  1.03  &  1.44  &  1.33  &  1.55  &  B  &   \\
 75   &  -- & -- &  1.37  &  1.37  &  1.38  &  1.64  &  1.35  &  1.50  &     &   \\
 76   &  -- & -- &  1.58  &  1.51  &  1.61  &  1.83  &  1.78  &  1.75  &     & M1  \\
 79   &  Y  & -- &   --   &   --   &   --   &   --   &   --   &   --   &  C  &   \\
 80   &  -- & -- &  1.27  &  1.39  &  1.29  &  0.96  &  0.92  &  0.98  &  B  &   \\
 81   &  -- & -- &  1.59  &  1.52  &  1.66  &  1.50  &  1.28  &  1.42  &     & M5  \\
 82   &  -- & -- &  1.31  &  1.23  &  1.27  &  1.14  &  0.90  &  1.15  &  E  & M0  \\
 83   &  -- & -- &   --   &   --   &   --   &   --   &   --   &   --   &  E  &   \\
 86   &  -- & -- &  1.80  &  1.61  &  1.83  &  1.28  &  0.98  &  1.23  &     & M5.5  \\
 87   &  -- & -- &   --   &   --   &   --   &   --   &   --   &   --   &  B  & M4  \\
 88   &  -- & -- &  2.58  &  2.75  &  2.53  &  1.70  &  1.41  &  1.47  &     & M0.5  \\
 89   &  -- & -- &  1.10  &  1.21  &  1.08  &  1.62  &  1.68  &  1.71  &  B  & K5  \\
 90   &  -- & -- &  1.19  &  1.19  &  1.21  &  1.54  &  1.62  &  1.67  &  B  &   \\
 91   &  Y  & -- &   --   &   --   &   --   &   --   &   --   &   --   &  E  & K7  \\
 92   &  -- & -- &  1.56  &  1.60  &  1.63  &  1.92  &  1.66  &  1.97  &  E  & M0  \\
 94   &  -- & -- &  1.14  &  0.85  &  0.99  &  0.46  &  0.53  &  0.54  &     &   \\
 96   &  -- & -- &  1.35  &  1.23  &  1.34  &  1.53  &  1.48  &  1.41  &  E  & M1  \\
 97   &  -- & Y  &   --   &   --   &   --   &   --   &   --   &   --   &     & M2  \\
 98   &  -- & Y  &   --   &   --   &   --   &   --   &   --   &   --   &     & A3  \\
100   &  -- & -- &  1.39  &  1.47  &  1.39  &  1.90  &  1.56  &  1.89  &     &   \\
101   &  -- & -- &  1.60  &  1.55  &  1.63  &  1.95  &  1.65  &  1.92  &  B  &   \\
102   &  Y  & -- &   --   &   --   &   --   &   --   &   --   &   --   &  B  &   \\
103   &  -- & -- &  1.27  &  1.53  &  1.31  &  1.30  &  1.28  &  1.16  &     &   \\
104   &  -- & -- &  1.50  &  1.67  &  1.53  &  1.96  &  1.70  &  1.90  &     &   \\
105   &  -- & -- &  0.87  &  0.87  &  0.80  &  3.84  &  3.15  &  3.30  &     &   \\
106   &  -- & -- &  1.19  &  1.28  &  1.19  &  1.22  &  1.42  &  1.61  &     & M3  \\
107   &  -- & -- &  1.07  &  1.04  &  1.06  &  1.24  &  1.16  &  1.37  &     &   \\
109   &  -- & -- &   --   &   --   &   --   &  1.58  &  1.31  &  1.51  &  B  &   \\
111   &  -- & -- &  1.24  &  1.32  &  1.29  &  0.61  &  1.11  &  1.23  &     &   \\
113   &  -- & -- &  1.10  &  1.25  &  1.12  &  1.76  &  1.82  &  1.91  &     & K7  \\
114   &  -- & -- &  1.14  &  1.21  &  1.16  &  1.90  &  1.99  &  2.05  &  B  & F9  \\
115   &  -- & -- &  1.43  &  1.36  &  1.45  &  2.22  &  2.06  &  2.11  &     & M0.5  \\
116   &  -- & -- &   --   &   --   &   --   &  1.56  &  1.51  &  1.49  &  B  &   \\
117   &  -- & -- &  1.37  &  1.38  &  1.40  &  1.87  &  1.68  &  1.81  &     & K2  \\
119   &  -- & -- &  1.20  &  1.23  &  1.25  &  1.22  &  0.82  &  1.05  &  B  & K7  \\
120   &  -- & Y  &   --   &   --   &   --   &  1.70  &  1.52  &  1.64  &  B  & A2  \\
122   &  -- & -- &  1.22  &  1.17  &  1.21  &  0.90  &  1.08  &  1.09  &  B  & M0  \\
123   &  -- & -- &  1.87  &  1.70  &  1.90  &  1.97  &  1.63  &  1.95  &  B  & M0  \\
124   &  -- & -- &  1.09  &  1.15  &  1.13  &  1.06  &  1.15  &  1.08  &  B  & M0  \\
125   &  -- & -- &  1.57  &  1.50  &  1.61  &  2.06  &  1.86  &  2.02  &  B  & M0  \\
127   &  -- & -- &   --   &   --   &   --   &  1.85  &  1.71  &  2.00  &  B  & M2  \\
129   &  -- & -- &  1.19  &  1.16  &  1.20  &  1.11  &  0.95  &  1.15  &     &   \\
130   &  -- & -- &   --   &   --   &   --   &  1.39  &  1.24  &  1.23  &     & K6  \\
131   &  -- & Y  &   --   &   --   &   --   &   --   &   --   &   --   &     & A3  \\
132   &  Y  & -- &   --   &   --   &   --   &   --   &   --   &   --   &  B  &   \\
133   &  Y  & -- &   --   &   --   &   --   &   --   &   --   &   --   &  B  &   \\
134   &  -- & -- &  0.76  &  0.88  &  0.73  &  1.25  &  1.33  &  0.89  &  B  &   \\
135   &  Y  & -- &   --   &   --   &   --   &   --   &   --   &   --   &  B  &   \\
136   &  -- & -- &  1.08  &  0.98  &  1.07  &  1.30  &  1.38  &  1.54  &  B  &   \\
137   &  -- & -- &  1.27  &  1.24  &  1.28  &  1.99  &  2.06  &  2.01  &  B  &   \\
139   &  -- & -- &   --   &   --   &   --   &  1.88  &  1.50  &  1.76  &     & A4  \\
140   &  Y  & -- &   --   &   --   &   --   &   --   &   --   &   --   &  B  &   \\
142   &  -- & -- &  1.50  &  1.39  &  1.49  &  1.26  &  1.13  &  1.16  &  B  & M4  \\
143   &  -- & -- &   --   &   --   &   --   &  1.42  &  1.26  &  1.47  &  B  &   \\
144   &  -- & -- &  0.98  &  1.11  &  0.99  &  1.30  &  1.23  &  1.32  &  B  &   \\
145   &  -- & -- &   --   &   --   &   --   &  1.27  &  1.26  &  1.32  &  B  & G2.5  \\
146   &  -- & -- &  1.23  &  1.39  &  1.25  &  1.28  &  1.15  &  1.22  &  A  & M4  \\
147   &  -- & -- &  1.41  &  1.51  &  1.46  &  1.40  &  1.23  &  1.42  &  B  &   \\
148   &  -- & -- &  1.81  &  1.67  &  1.83  &  1.76  &  1.51  &  1.79  &  B  & K7  \\
149   &  -- & -- &  1.90  &  1.65  &  1.89  &  1.47  &  1.31  &  1.43  &     & M0  \\
\enddata
\tablenotetext{a}{All disk sources with PAH in emission show the
  features at 6.2, 7.7, 8.6, 11.2 and 12.8 $\mu$m }
\end{deluxetable}

\begin{deluxetable}{r c c c }
\tablecolumns{4}
\tablewidth{0pt}
\tablecaption{Number of Objects in Tight Groups \& Associations \label{tcluster}}
\tablehead{\colhead{ }              &
           \colhead{Embedded}       &
           \colhead{Disks}          &
           \colhead{Total} 
}
\startdata
Cluster A &  0  &  1   &   1\tablenotemark{a} \\
Cluster B &  5  & 26   &  31                  \\
Group C   &  9  & 10   &  19                  \\
Group D   &  0  &  8   &   8                  \\
Group E   &  3  & 12   &  15                  \\
\enddata
\tablenotetext{a}{The Serpens Core, or Cluster A, is not complete in
  these observations.}
\end{deluxetable}

\begin{table}
\caption{Statistics of the YSO populations in Serpens and Taurus }
\begin{center}
\begin{tabular}{l c | c c c c }
\hline
\hline
 & Embedded  &  \multicolumn{4}{c}{Disks} \\
 &           &  PAH  &  both 10 and 20 $\mu$m  &  only 20 $\mu$m  &  featureless \\
\hline
Serpens &  18.3\%  &   2.6\%   &  60.0\%  &  13.9\%  &   4.3\%  \\
Taurus  &  20.1\%  &   2.9\%   &  57.6\%  &   3.6\%  &  18.7\%  \\
\hline
\end{tabular}
\end{center}
\label{tstats}
\end{table}

\clearpage

\appendix

\section{Background Sources}
\label{abg}

Here the background contamination discussed in \S\ref{sbg} is
presented in 3 categories.

\subsection{Background stars}
\label{sbgs}

The IRS spectra of bright background stars, when seen through a
molecular cloud, show the silicate feature at 10 $\mu$m in absorption
and not in emission as for disk sources (see \S~\ref{ssi}), on top of a
falling infrared spectrum. \citet{OL09} identified candidate
background stars based on optical spectra for 78 objects in this
sample. After determining spectral types, extinctions and
luminosities, those objects were placed in a Hertzsprung-Russell
diagram (HRD) and compared with 2 sets of isochrones and mass tracks
(\citealt{BA98}, \citealt{SI00}). 20 out of those 78 objects proved to
be much too luminous to be at the distance of Serpens ($d = 259 \pm
13$pc, \citealt{ST96}). Those objects were then classified as
background sources, typically asymptotic giant branch (AGB) stars with
dusty shells. Interestingly, 18 of the 20 objects classified as
background in \citet{OL09} have IRS spectra characteristic of
background sources; the other two objects, \#81 and 86, show silicates
in emission (just like those seen in circumstellar disks) and were
then re-classified as YSOs.

\begin{figure}[ht!]
\begin{center}
\includegraphics[width=\textwidth]{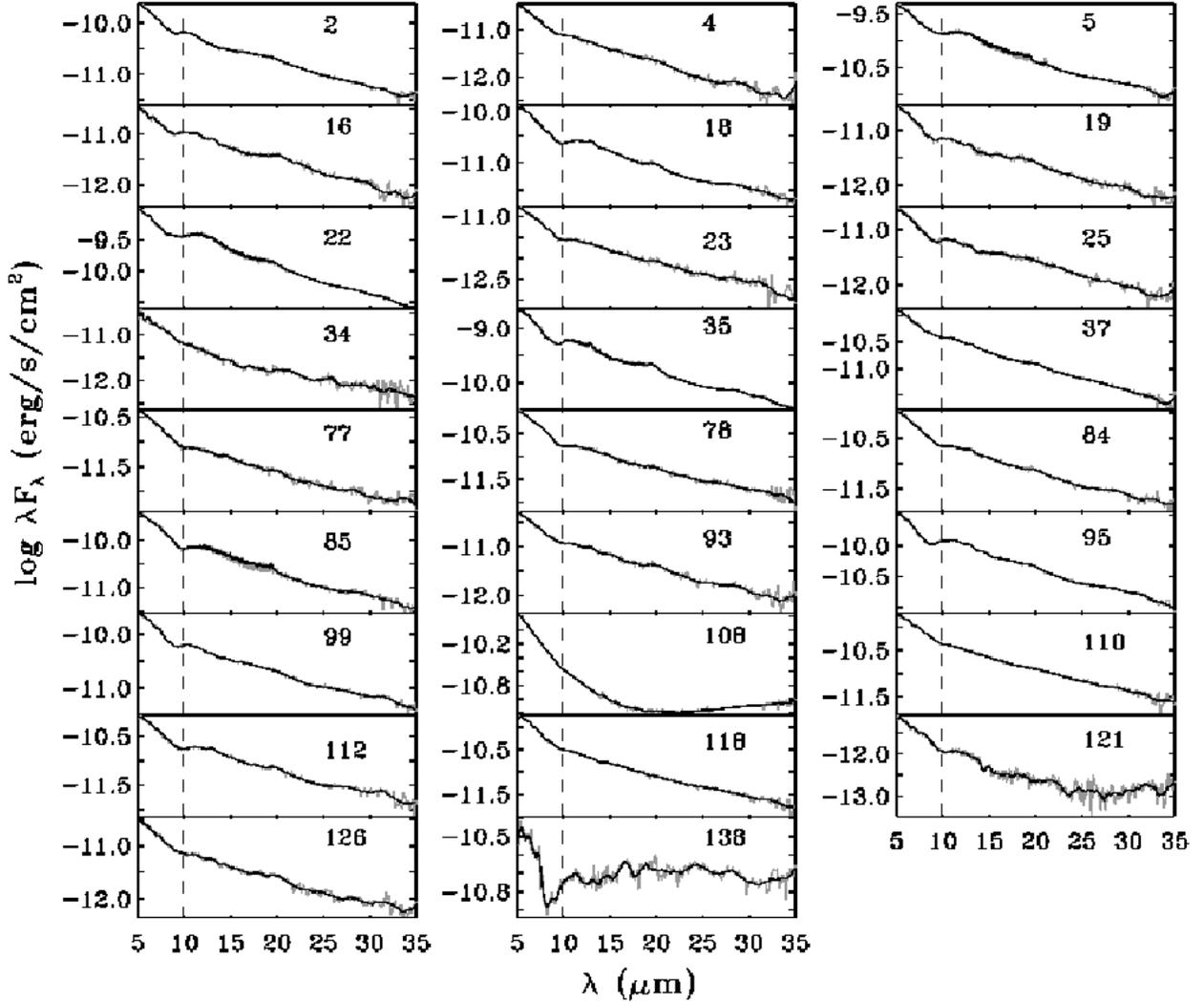}
\end{center}
\caption{\label{irsbg} IRS spectra of the background stars toward
  Serpens with the characteristic 10 $\mu$m silicate absorption
  feature and falling SED. In grey are the original spectra, while in
  black a binned version is overplotted. The object \# is indicated at
  the top right of each panel. }
\end{figure}

Besides these 18 objects, another 8 objects in this sample with
falling mid-IR spectra show the 10 $\mu$m silicate feature in
absorption (objects \#19, 25, 37, 78, 95, 126 were not observed by
\citealt{OL09}), leading to their classification as background
stars. Alternatively, those 8 objects could be highly extincted YSOs,
with silicate features in absorption arising from such
extinction. However, using the dense cloud dust properties, as in
\S~\ref{ssi_ext}, an extinction of 10 mag would produce a $\tau_{9.7}$
of only 0.14 (see also \citealt{CH07}). Since these objects do have
silicate features in absorption, it is understood that they have high
extinction (considerably higher than what is found for the cloud at
their location). Thus, it is unlikely that they are actually highly
extincted cloud members. Our working assumption is that these are
background objects. These 26 spectra are shown in Figure \ref{irsbg}.

\subsection{Background galaxies}
\label{sbgg}

Five of the sources with clear PAH features show wavelength shifts in
their line positions compared with galactic sources. These features
thus allow us to identify them as background galaxies in this
sample. PAHs are common features in most nearby galaxies with ongoing
or recent star-formation \citep{SM07}, and some contamination by
background galaxies is expected \citep{PP04}.

\begin{figure}[h!]
\begin{center}
\includegraphics[width=0.5\textwidth]{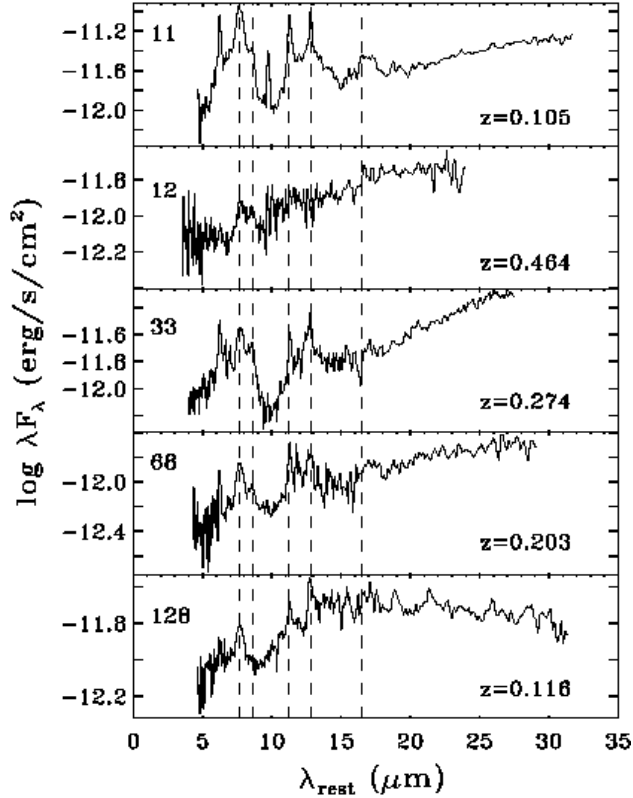}
\end{center}
\caption{\label{irsgal} IRS spectra of the background galaxies toward
  Serpens, shown in rest-frame wavelength. The dashed lines indicate
  the rest-frame PAH features at 6.2, 7.7, 8.6, 11.2, 12.8 and 16.4
  $\mu$m. The object \# and $z$ are indicated in each panel.}
\end{figure}

By identifying PAH features from their shape and pattern
(\citealt{TI08} and references therein), the observed wavelength
($\lambda_{\mathrm{obs}}$) can be related to the rest-frame wavelength
($\lambda_{\mathrm{rest}}$) of each feature through a redshift ($z =
\frac{\lambda_{\mathrm{obs}}}{\lambda_{\mathrm{rest}}} -
1.0$). Between 3 (\#12) and 5 (\#11 and 68) PAH features are
identified in each spectrum, and the redshift of each line is
calculated. For a given object, the $z$ values derived from the
various lines agree within 2\% and the mean of these values is taken
as the redshift for that source. These 5 redshifted galaxies can be
seen, in rest-frame wavelength, in Figure \ref{irsgal}.

Star-forming galaxies show strong correlations between the
extinction-corrected Paschen $\alpha$ luminosity coming from HII
regions and the luminosity at 24 $\mu$m due to the emission of dust
heated by hot, young stars. \citet{RI09} obtained power-law fits to
the relation between the star formation rate (SFR) and the observed 24
$\mu$m flux densities ($f_{24,obs}$), parameterized with respect to
$z$ to take into account the fact that the rest-frame wavelengths
sampled change with redshift (Eq. 14 in their paper). In order to
estimate the infrared luminosity ($L(\rm{TIR})$), we use their
best-fit relations between the (rest-frame) 24 $\mu$m luminosity,
$L(\rm{TIR})$ and SFR (Eqs. 11 and 25).

We have calculated SFRs\footnote{The SFRs correspond to an IMF
  according to \citet{GC03}.} and $L$(TIR) for the 5 background
galaxies contaminating this sample. These results can be found in
Table \ref{tgal}. $L(\rm{TIR})$ ranges from a few times 10$^{11}$ to a
few times 10$^{12}$ L$_\odot$, indicating that these galaxies are
typical so-called luminous infrared galaxies (LIRGs,
$L(\rm{TIR})>10^{11}$ $L_\odot$), while two are ultra-luminous
(ULIRGs, $L(\rm{TIR})>10^{12}$ $L_\odot$). The latter are a class of
heavily obscured galaxies having enormous instantaneous SFRs that are
often associated with merging galaxies \citep{SM96}. These galaxies
are very rare locally and become more common at high redshift. Their
space density rises by about a factor of 10 from $z=0$ to $z\sim0.5$
\citep{MA09}, consistent with the relatively high redshifts found
($z\sim0.3-0.5$). Furthermore, due to its high SFR and the empirical
correlation between the infrared and radio emission from starburst
galaxies \citep{CO92}, the ULIRG \#12 at $z=0.46$ can be identified
with a previously unidentified radio source in the NRAO VLA Sky Survey
(NVSS J182817+001604, \citealt{CO98}).

The inferred SFRs range between 11 and 731 M$_\odot$ yr$^{-1}$,
compared to a Milky Way SFR of about 3--5 M$_\odot$ yr$^{-1}$
(\citealt{PA95}, and references therein). For further reference, the
SFR of Serpens is 5.7$ \times 10^{-5}$ M$_\odot$ yr$^{-1}$
\citep{EV09}. Even though the star formation observed in those
galaxies is due to high mass star-forming regions such as Orion, the
high SFRs in the LIRGs and ULIRGs found are equivalent to $10^5 -
10^7$ molecular clouds such as Serpens!

\subsection{High Ionization Object}
\label{sbgpn}

\begin{figure}[h!]
\begin{center}
\includegraphics[width=0.6\textwidth]{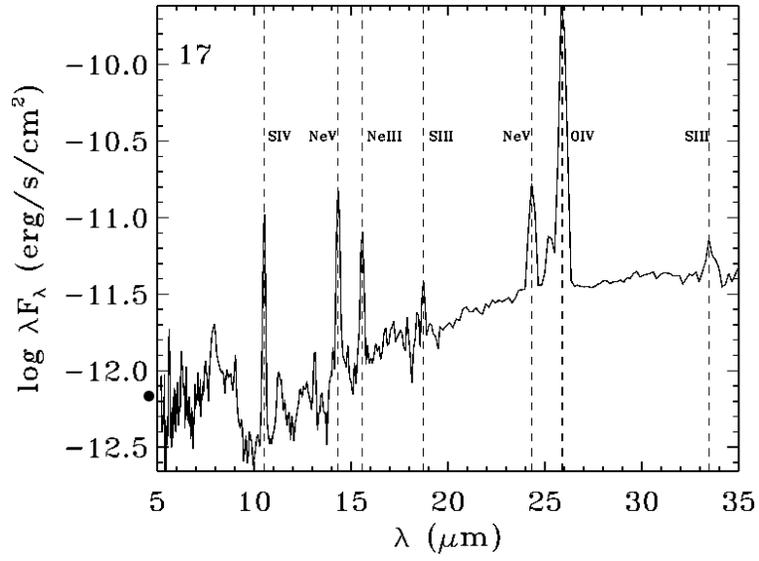}
\end{center}
\caption{\label{irspn} IRS spectra of the planetary nebula
  candidate(\#17) found among this sample, presenting high ionization
  lines. }
\end{figure}

One of the objects, \#17 (SSTc2dJ18282720+0044450), in this sample
presents very strong emission lines that do not seem compatible with
YSOs.  Figure \ref{irspn} shows a blow-up of the spectrum, identifying
the high ionization lines at $z=0$. The source itself is highly
reddened, being barely detected in IRAC1 (3.5 $\mu$m) but very
prominent in MIPS1 (24 $\mu$m), with a 3--24 $\mu$m slope of 1.82. Two
types of galactic objects show such high excitation lines: dusty
planetary nebulae and supernova remnants. The IRS spectrum of this
object (\#17) resembles those of shocked ejecta of supernova remnants,
as in \citet{SA09} and \citet{GA09} and that of planetary nebulae such
as LMC 78, shown in \citet{BS09}.

The spectrum also shows PAH emission at 6.2, 7.7, 11.2, and 12.8
$\mu$m. Optical and near-IR spectra are necessary to confirm the
nature of this object.

\clearpage

\begin{deluxetable}{r c c c c c c }
\tabletypesize{\footnotesize}
\tablecolumns{7}
\tablewidth{0pt}
\tablecaption{Characteristics of the Background Galaxies \label{tgal}} 
\tablehead{\colhead{\#}                       & 
           \colhead{$z$}                      &
           \colhead{24$\mu$m (Jy)}            &
           \colhead{SFR (M$_\odot$ yr$^{-1}$)}  &
           \colhead{$L(\rm{TIR})$ (L$_\odot$)}      &
           \colhead{Type}     
}
\startdata
11                    &  0.1051  &  0.01800  &   16  &  1.5$ \times 10^{11}$  &  LIRG   \\
 12\tablenotemark{a}  &  0.4637  &  0.00893  &  731  &  4.8$ \times 10^{12}$  &  ULIRG  \\
 33                   &  0.2742  &  0.01490  &  217  &  1.6$ \times 10^{12}$  &  ULIRG  \\
 68                   &  0.2032  &  0.00909  &   47  &  4.0$ \times 10^{11}$  &  LIRG   \\
128                   &  0.1164  &  0.01010  &   11  &  1.1$ \times 10^{11}$  &  LIRG   \\
\enddata
\tablenotetext{a}{NVSS J182817+001604}
\end{deluxetable}

\end{document}